\documentclass[sigplan,10pt]{acmart}

\usepackage{hyperref}
\usepackage[small,compact]{titlesec}
\usepackage{mathtools}
\usepackage{multirow}
\usepackage{wrapfig}
\usepackage{url}
\usepackage{xcolor}
\usepackage{enumitem}
\usepackage{capt-of}
\usepackage{booktabs}
\usepackage{subcaption}
\usepackage{calrsfs}
\usepackage[linesnumbered,noend]{algorithm2e}
\usepackage{etoolbox}% <---- needed for this solution
\providetoggle{beginFlag}
\settoggle{beginFlag}{true}

\SetAlgoSkip{SkipBeforeAndAfter}

\usepackage{xspace}
\usepackage{soul}
\usepackage[noabbrev]{cleveref}
\newcommand{\mli}[1]{\mathit{#1}}

\newcommand{\eg}{{\it e.g.}, }
\newcommand{\ie}{{\it i.e.}, }

\newcommand{\system}{\textsc{Bagpipe}\xspace}
\newcommand{\fae}{\textsc{FAE}\xspace}
\newcommand{\torchrec}{\textsc{TorchRec}\xspace}
\newcommand{\het}{\textsc{HET}\xspace}
\newcommand{\embs}{Embedding Server\xspace}
\newcommand{\orc}{Oracle Cacher\xspace}
\newcommand{\lkval}{$\mathcal{L}$\xspace}

\newenvironment{myitemizeleft}
{
   \vspace{0pt}
    \begin{list}{$\bullet$ }{\leftmargin=1em \itemindent=0em}
        \setlength{\topsep}{0em}
        \setlength{\parskip}{0pt}
        \setlength{\partopsep}{0pt}
        \setlength{\parsep}{0pt}
        \setlength{\itemsep}{0.3mm}
}
{
    \end{list}
}

\usepackage{graphicx}
\graphicspath{ {./images/} }

\begin{document}
\pagenumbering{gobble}
\title{\system: Accelerating Deep Recommendation Model Training}
\author{Saurabh Agarwal}
\affiliation{
\institution{University of Wisconsin-Madison}
\country{}
}
\author{Chengpo Yan}
\affiliation{
\institution{University of Wisconsin-Madison}
\country{}
}
\author{Ziyi Zhang}
\affiliation{
\institution{University of Chicago}
\country{}
}
\author{Shivaram Venkataraman}
\affiliation{
\institution{University of Wisconsin-Madison}
\country{}
}
%\author{
%{\rm Your N.\ Here}\\
%Your Institution
%\and
%{\rm Second Name}\\
%Second Institution
% copy the following lines to add more authors
% \and
% {\rm Name}\\
%Name Institution
% \author{Paper \# 175} % end author
\begin{abstract}
Deep learning based recommendation models (DLRM) are widely used in several business critical applications. Training such recommendation models efficiently is challenging because they contain billions of embedding-based parameters, leading to significant overheads from embedding access. By profiling existing systems for DLRM training, we observe that around 75\% of the iteration time is spent on embedding access and model synchronization. Our key insight in this paper is that embedding access has a specific structure which can be used to accelerate training. We observe that embedding accesses are heavily skewed, with around 1\% of embeddings representing more than 92\% of total accesses. Further, we also observe that during offline training we can lookahead at future batches to determine which embeddings will be needed at what iteration in the future. Based on these insights, we develop \system{}, a system for training deep recommendation models that uses caching and prefetching to overlap remote embedding accesses with the computation. We design an Oracle Cacher, a new component that uses a lookahead algorithm to generate optimal cache update decisions while providing strong consistency guarantees against staleness. We also design a logically replicated, physically partitioned cache and show that our design can reduce synchronization overheads in a distributed setting. Finally, we propose a disaggregated system architecture and show that our design can enable low-overhead fault tolerance. Our experiments using three datasets and four models show that \system{} provides a speed up of up to 5.6x compared to state of the art baselines, while providing the same convergence and reproducibility guarantees as synchronous training.

% which are often stored remotely
% is spent in forward/backward pass while the remaining time is

% Finally, we show that all optimizations introduced by Bagpipe are transparent to the user and thus provide the same

\end{abstract}



% \date{}
%\date{University of Wisconsin-Madison}
% \settopmatter{printfolios=true}
\acmYear{2023}\copyrightyear{2023}
\setcopyright{acmlicensed}
\acmConference[SOSP '23]{ACM SIGOPS 29th Symposium on Operating Systems Principles}{October 23--26, 2023}{Koblenz, Germany}
\acmBooktitle{ACM SIGOPS 29th Symposium on Operating Systems Principles (SOSP '23), October 23--26, 2023, Koblenz, Germany}
\acmPrice{15.00}
\acmDOI{10.1145/3600006.3613142}
\acmISBN{979-8-4007-0229-7/23/10}

\maketitle
% \printlength\textwidth
\pagestyle{plain}

% \section{TODO Measurements}
% \subsection{Motivation and Intro}
% \begin{enumerate}
%     \item Plot Size of embedding tables
%     \item Plot Embedding access and communication time per iteration for few different configurations. \todo{What models to use. We can create a compute heavy model as well. The facebook paper doesn't go in detail of different type of models.}
%     \item hotness changes \newline

%     \item Plot portion of cache hit within batches. When different \% of embeddings are stored. 
%     \item Can we get something like Figure 3 in \cite{adnan2021high} which shows that as we increase the batch size we will not get batches with hot embeddings.

% \end{enumerate}

% \vspace{-5pt}
\section{Introduction}
% \vspace{-5pt}
% suggest friends and personalise content, e-commerce websites use them to recommend products. 
Recommendation models are widely deployed in enterprises to personalize and improve user experience. Applications that use recommendations range from personalized web search results~\cite{su2017improving} to friend recommendations in social networks~\cite{liu2017related,deng2016deep} and product recommendations in e-commerce \cite{li2020improving}. 
With growing data sizes~\cite{acun2021understanding} and the use of the recommendation in increasingly sophisticated tasks~\cite{he2017neural, lee2018collaborative}, recent trends have seen the adoption of new deep learning based recommendation models~\cite{cheng2016wide,dlrmopensource}. 
Deep learning based recommendation  models have become one of the largest ML workloads with companies like Meta reporting that
more than 50\% of all ML training cycles~\cite{acun2021understanding} and more than 70\% of
inference cycles~\cite{gupta2020deeprecsys} are being used for deep learning based recommendation (DLRM) models.

\begin{figure*}[t]
\begin{minipage}{0.32\textwidth}
    \vspace{-10pt}
    \begin{center}
    \includegraphics[width=0.85\textwidth]{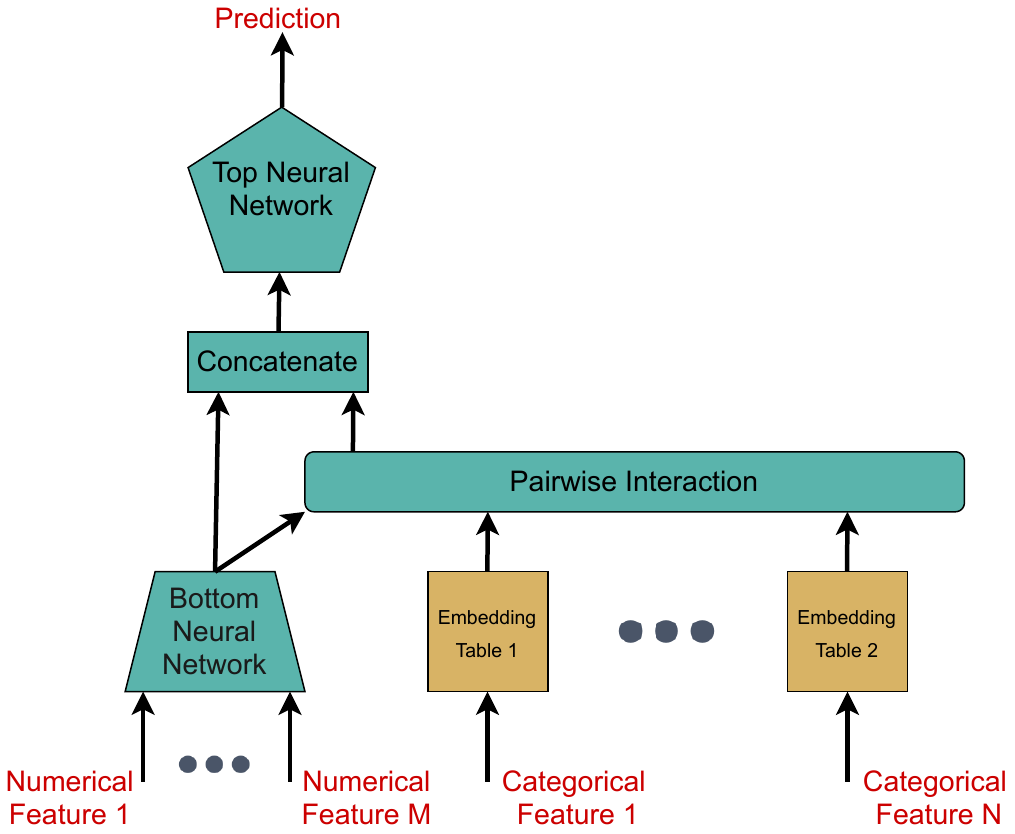}
    \vspace{-0.1in}
    \caption{\small{\textbf{Architecture of a recommendation model}: Model parameters include top, bottom NNs and embedding tables.}}   
    \label{fig:model_arch}
    \end{center}
\end{minipage}\quad
\begin{minipage}{0.65\textwidth}
\begin{center}
    \includegraphics[width=0.8\linewidth]{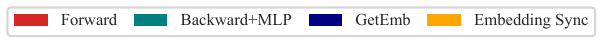}\\
    \vspace{-5pt}
    \end{center}
\begin{subfigure}{0.45\linewidth}
    % \vspace{-10pt}
    \begin{center}
    \includegraphics[width=\linewidth]{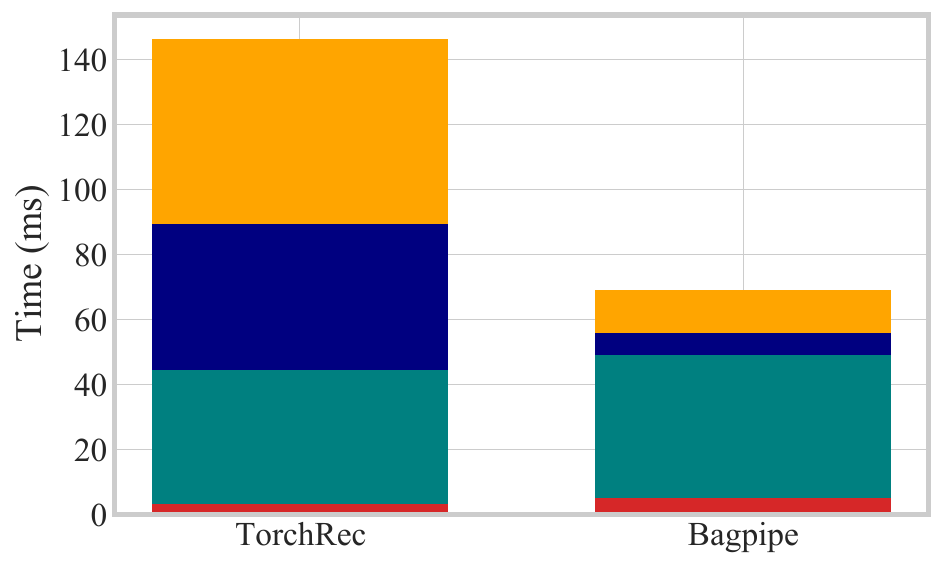}
    \vspace{-20pt}
    \caption{\small{DLRM}}
    \label{fig:time_distribution_dlrm}
    \end{center}
    \end{subfigure}
\begin{subfigure}{0.45\linewidth}
    % \vspace{-10pt}
    \begin{center}
    \includegraphics[width=\textwidth]{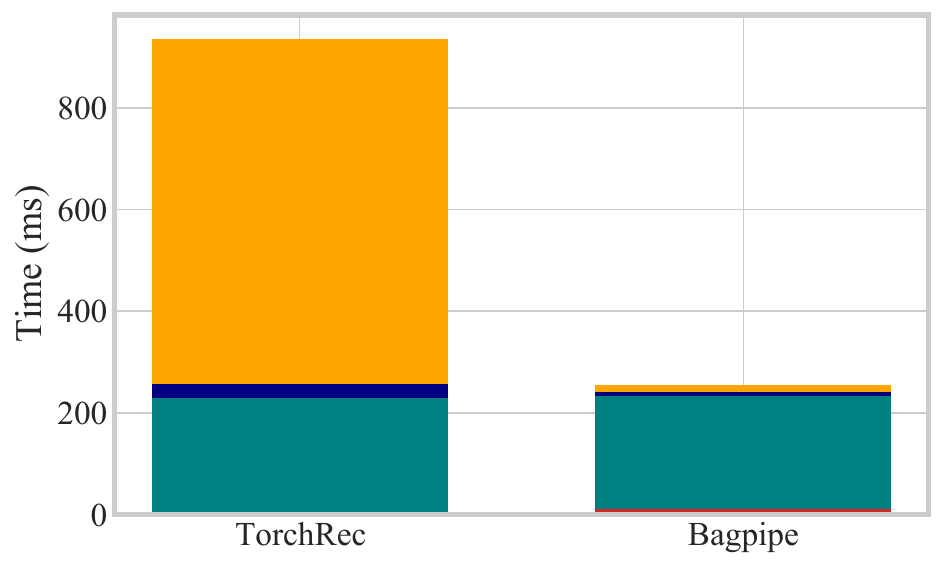}
    \vspace{-20pt}
    \caption{\small{DeepFM}}
    \end{center}
    \end{subfigure}
    \vspace{-0.1in}
    \caption{\small{\textbf{Training Time Breakdown:} Average time spent in various stages of training when using 8 \emph{p3.2xlarge} instances with \torchrec~\cite{torchrec} (left) and \system (right) on DLRM and DeepFM models (Table~\ref{tab:models}). For large models like DeepFM, we observe that \torchrec spends 75\% of each iteration on embedding access, while \system can bring it down to 10\%.
    % \shivaram{legend is broken. Shouldn't say Oboe. should say ideal. Also arrange all the torchrec legends in one row at the top.} 
    }}
    \label{fig:time_distribution}
\end{minipage}
\vspace{-10pt}
\end{figure*}

The structure of recommendation models differs significantly from other popular deep neural networks.
% 
%introducing new challenges to efficiently scale training. 
Recommendation models contain two types of model parameters: embedding tables that store vector representation of categorical features and neural networks that consist of multi-layer perceptrons (MLP) for numeric features (Figure~\ref{fig:model_arch}). If we consider a click-through-rate (CTR) prediction model, given a batch of training examples (user clicks), in the forward pass we first look up the relevant embeddings for categorical features of the input. We note that the embedding lookup is \emph{sparse}; for instance, if the user location is New York, we only need to fetch the embeddings corresponding to that location from the embedding table which contains embeddings for all the locations. The numerical features are processed by a \emph{dense} MLP, and the representations are then combined to form a prediction as shown in Figure~\ref{fig:model_arch}. Similar to existing DNNs, the embeddings and the MLP parameters are updated in the backward pass based on the gradients.  

% (bottom and top)

%During training each of the machines fetch relevant embeddings before forward pass, during backward pass the dense parameters as well as the embeddings are synchronized.
% \shivaram{we say several challenges but only highlight one?}
The unique structure of recommendation models introduces a new design challenge which is
not handled by existing DL training systems~\cite{narayanan2019pipedream,jia2019beyond,li2014scaling}.
The challenge arises from the extremely large, memory-intensive embedding tables. In
production systems it is common for size of embedding tables to be in terabytes~\cite{yin2021ttrec,
gupta2020deeprecsys} and Table~\ref{tab:emb_size} shows the embedding table sizes for three open-source datasets.
Since embedding tables will not fit in the memory of a single worker, a hybrid parallelization
approach~\cite{mudigere2021software, acun2021understanding, torchrec} is currently used to train these models.
With hybrid parallelization, dense MLP layers are \emph{replicated} across GPUs on each machine while
embedding tables are \emph{partitioned} and stored on the CPU memory of workers, leading to a design
where model parallelism is used for embeddings, and data parallelism is used for dense layers.
% layers MLP
% extremely large size

Despite using hybrid parallelism, existing approaches to DLRM training suffer from embedding access bottlenecks as each iteration of training requires remote access of embeddings. In Figure~\ref{fig:time_distribution}, profiling TorchRec~\cite{torchrec} on an 8 GPU cluster with the Criteo dataset (details in \S\ref{sec:eval}) shows that on each iteration, up to 75\% of the time is spent in embedding fetch and write back operations. Reducing the embedding access overhead forms the primary motivation of our work. 
% This motivates us to ask: \emph{can we design a system for training recommendation models that can alleviate embedding access bottlenecks?}

% of updated embeddings takes around 75\% of the overall iteration time

To reduce embedding access overhead we observe that the embedding accesses have a specific structure which can be leveraged to accelerate training.
%recommendation model  training has structur
%\shivaram{Removing pointer to figure to keep it brief} and patterns
Our analysis across three datasets indicates significant \emph{skew} in embedding accesses; \eg with the Kaggle (Criteo) dataset, we observe that 90\% of accesses are for just 0.1\% of the embeddings. While this might indicate that we can just \emph{cache} popular embeddings in GPU memory~\cite{adnan2021high}, we find that to be insufficient as each training batch also requires a number of embeddings that are not present in the cache. %  We find tha
When caching 0.1\% of the most popular embeddings for the same Criteo datasets with a batch size of 16,384, 
% We observe that for the same Criteo dataset with a batch size of 16,384 when caching 0.1\% of the most popular embeddings, 
only 15\% of the total unique embedding access are served from the cache(\S\ref{sec:access_patterns}).

Given the above embedding access patterns, to alleviate overheads, our key
insight is that when performing offline training of ML models we can \emph{lookahead} beyond the current batch and observe the data access of future batches. Thus, we can revisit classic perfect-caching algorithms~\cite{belady1966study} to \emph{prefetch} embeddings based on when they will be used and cache embeddings in GPU memory if they %are going to be 
%  as we can determine exactly which embeddings will be needed and at what iteration in the future. 
will be reused in the near future. Unlike prior systems, we propose jointly using prefetching and caching to overlap embedding fetches for future batches with compute of the current batch, effectively hiding the latency of embedding access. 
%and optionally .

% \shivaram{Splitting paragraph to separate contributions}
However, there are additional challenges in extending this lookahead-based design to a distributed setting. To ensure embeddings are not stale, we need to perform cache synchronization across workers which leads to
additional communication overheads.
We propose a new \emph{logically replicated, physically partitioned} (LRPP) cache design (\S\ref{sec:distributed_cache}) to minimize the cache synchronization time; having a logically replicated cache minimizes the overhead of tracking state separately for each trainer,
% in routing examples to trainers, 
while having a physically partitioned design helps in reducing the number of bytes transferred across trainers. We also enhance this design using Critical Path Analysis (CPA)~\cite{yang1988critical} to only synchronize necessary embeddings on the critical path and delay the rest, thereby overlapping part of cache synchronization with forward/backward compute of future iterations. In combination, we find that our techniques can reduce communication overheads 
% in every iteration 
by 65\%--70\%.
%cache contents of each trainer can vary  and 

%Based on these insights 
We build the above techniques in \system, a system for large-scale distributed training of recommendation models.
% \system is designed to achieve three primary goals- (i) alleviate embedding access overhead, (ii) provide same semantics and guarantees as synchronous training and (iii) allow independent scaling of memory intensive (Embedding servers) and compute intensive (trainers) components based on workload.
% propose a disaggregated architecture that separates trainers from . 
% \system's disaggregated architecture separates trainers from  embedding storage (\S\ref{sec:system_design}) which allows independent scaling of compute intensive components and memory intensive components based on model and dataset requirements. 
Central to our design is an \emph{Oracle Cacher}, a new service that looks beyond the current batch, to 
determine which embeddings to pre-fetch and/or cache. 
The caching decisions made by \orc are realized on each training worker, and trainers overlap prefetching and cache synchronization with model training. 
We show that this design enables independent scaling of various training components based on workload requirements and also minimizes the time required to recover from failures 
(\S\ref{sec:ft}).

To evaluate \system we use three datasets, Criteo Kaggle~\cite{kaggle}, Avazu~\cite{avazu}, and Criteo Terabyte~\cite{terabyte}, and four popular models, DLRM~\cite{naumov2019deep}, Wide\&Deep~\cite{cheng2016wide}, DeepFM~\cite{guo2017deepfm} and D\&C~\cite{wang2017deep}. 
%\shivaram{update after S5 is complete}
We scale up our training to model sizes of 4.4 billion parameters, (similar to models used in MLperf) using up to 32 GPUs. Overall we find that \system can improve iteration time by $3.7\times$ compared to \torchrec~\cite{torchrec} and $2.3\times$ compared to asynchronous training in HET~\cite{miao2021het}.  We also show that \system's design enables fault tolerance with low overhead and can recover from a trainer failure 13$\times$ faster compared to FB-Research's system~\cite{dlrmopensource}.  
Finally, unlike prior work~\cite{adnan2021high,miao2021het}, our optimizations in \system{} maintain consistent access to embeddings. Thus, we maintain the same statistical efficiency as synchronous training, and our time-per-iteration improvements directly translate to 
% end-to-end 
time-to-accuracy speedups.

By using lookahead to fetch parts of the model (embeddings) out-of order and LRPP distributed caches,  
% to the best of our knowledge,
\system is the first distributed recommendation model training system that: (i) alleviates embedding access overhead by up to 92\% using both prefetching and caching, (ii) transparently accelerates training while maintaining the same guarantees as synchronous training and 
(iii) reduces network overheads and thus enables an efficient, dis-aggregated deployment where we can independently scale memory intensive (embedding servers) and compute intensive (trainers) workers based on requirements.

 % \vspace{-5pt}
\section{Background \& Motivation}
 % \vspace{-5pt}
We first provide background on recommendation model training and then motivate the need for a new system to optimize data movement.  

% \subsection{Background}

 % \vspace{-5pt}
\subsection{Deep Learning Recommendation Models} 
 % \vspace{-5pt}
Recommendation models power widely used large-scale internet services.
% are widely used in industry to power several large-scale internet services. 
Recently deep learning is being used to improve the accuracy of recommendations~\cite{naumov2019deep,cheng2016wide,ishkhanov2020time}. All deep recommendation models consist of two types of components (i) a memory-intensive embedding layer that stores a mapping between the categorical features and their numerical representations (ii) a compute-intensive neural network-based modeling layer which models interactions between numerical features and vector representations of categorical features. Across models, the structure of embedding tables typically remains the same. The number of rows (elements in the table) usually depends on the dataset, \ie the number of categories, while machine learning engineers vary the dimension of embedding \ie the size of the vector to represent a categorical feature. Common dimensions of embeddings are 16, 32, 48, and 64 but sometimes can be as large as 384~\cite{mudigere2021software}. However, the rows in  embedding tables vary widely and can be as small as 3 elements or as big as a few billion elements depending on the dataset~\cite{yin2021ttrec,gupta2020deeprecsys}. Neural network layers have more diversity, and the type of neural network usually depends on the modeling task at hand. DLRM~\cite{naumov2019deep}, a recommendation model popular at Meta 
% used for predicting click-through-rate, 
uses fully connected layers for both bottom and top neural networks.
While, DeepFM~\cite{guo2017deepfm} uses a factorization module that learns up to two-order feature interactions between sparse and dense features (Table~\ref{tab:models} summarizes other models we consider in this paper).
%while TBSM~\cite{ishkhanov2020time}, a model used for predicting a user's next interaction based on previous interactions uses a modified version of multi-head attention as the top neural network while still using a fully connected neural network for bottom neural network. 
The forward pass of training involves looking up embedding vectors corresponding to the data items. All deep learning based recommendation models ~\cite{wang2017your,wen2018visual, wang2014improving, van2013deep, covington2016deep} use this step to handle categorical features such as location, product type, gender, etc. 
Our focus is to reduce data access overheads that arise from performing embedding lookups and thus speed up the training of all recommendation models which use embedding tables.
\begin{table}[!t]
\caption{\small{Dataset and their embedding tables}}
\vspace{-5pt}
\label{tab:datasets}
\label{tab:emb_size}
\resizebox{\linewidth}{!}{
\begin{tabular}{@{}|l|lll|lll|@{}}
\toprule
              & \multicolumn{3}{c|}{Training Input}    & \multicolumn{3}{c|}{Embedding Tables}                                                                                           \\ \cmidrule(l){1-7} 
Datasets      & Datapoints   & \begin{tabular}[c]{@{}l@{}}Categorical \\ Features\end{tabular} & \begin{tabular}[c]{@{}l@{}}Num \\ Features\end{tabular}          & \begin{tabular}[c]{@{}l@{}}Num\\ Emb\end{tabular} & \begin{tabular}[c]{@{}l@{}}Embedding \\ Dimension\end{tabular}  & \begin{tabular}[c]{@{}l@{}}Table \\Size \end{tabular}\\ \midrule
Kaggle Criteo & 39.2 Million & 26                                                              & 13                                                              & 33.76 Million                                                  & 48                         & 6 GB                                    \\
Avazu         & 40.4 Million & 21                                                              & 1                                                            & 9.4 Million                                                    & 48                       & 1.7 GB                                      \\
Terabyte      & 4.37 Billion & 26                                                              & 13                                                            &   882.77 Million                                                 & 16                       & 157 GB                                      \\ \bottomrule
\end{tabular}}
\vspace{-15pt}
\end{table}

\begin{table}[t]
\begin{center}
 \caption{\small{\textbf{Model Descriptions:} For DLRM, W\&D, D\&C the numbers indicate the structure of the different Fully Connected (FC) layers. For DeepFM, Linear Features represent a linear layer that is used to store feature interactions. For all the models, we used the standard architectures as suggested by original authors.}}
\vspace{-5pt}
\label{tab:models}
\resizebox{0.8\linewidth}{!}{
\begin{tabular}{@{}|l|l|l|@{}}
\toprule
Model  & Architecture of Dense Parameters                                                                             & \begin{tabular}[c]{@{}l@{}}Number of \\ Dense Parameters\end{tabular} \\ \midrule
DLRM~\cite{naumov2019deep}   & \begin{tabular}[c]{@{}l@{}}FC - 13-512- 256-64-48 \\ FC - 1024-1024-1024-256-128-1\end{tabular} & 2962289                                                               \\ \midrule
W\&D~\cite{cheng2016wide}   & FC - 13-256-256-256                                                                                  & 136673                                                                \\ \midrule
D\&C~\cite{wang2017deep}   & \begin{tabular}[c]{@{}l@{}}FC - 1024-512-256-64-48\\ FC - 1024-512-256-1\end{tabular}         & 2718609                                                               \\ \midrule
DeepFM~\cite{guo2017deepfm} & \begin{tabular}[c]{@{}l@{}}Linear Features - 33762577-1\\ FC - 1248-64-64-64\end{tabular}                   & 33851283                                                              \\ \bottomrule
\end{tabular}}
\end{center}
\vspace{-10pt}
\end{table}

\subsection{Training Recommendation models}
% \vspace{-5pt}
\label{sec:existing_method}
Next, we discuss the state-of-the-art systems used for training recommendation models.

\noindent\textbf{Offline Training vs Online Training.} Recommendation models are trained in both online and offline mode. 
% Offline training is a crucial component of DLRM training workloads, and can be viewed as pre-training, where a model is trained on large amounts of historical data and is throughput sensitive. 
Offline training %is a big part of DLRM training workloads, 
involves training the model on large amounts of historical data with emphasis on throughput. Alternatively, 
online training is performed only on the recently acquired data to account for latest user preferences and is latency sensitive. 
To boost performance and prevent catastrophic forgetting~\cite{french1999catastrophic,kumaran2016learning}, researchers actively perform offline training, even for models in production. According to a study by Meta~\cite{acun2021understanding} offline training is responsible for more than 50\% cycles of all ML model training cycles.
This shows that offline training of recommendation models is an important workload. In this work our primary focus is on offline training of recommendation models. 

\noindent\textbf{Training Setup.}
Recommendation models are extremely large and are currently among the largest ML models used in enterprises. Meta recently released a 12 Trillion parameter recommendation model~\cite{mudigere2021software}; in comparison GPT-3 has 175 Billion parameters. However, embedding tables with sparse access patterns account for more than 99\% of parameters.
% more than 99\% of parameters in these models are in embedding tables which have sparse access patterns. 
The combination of extremely large model sizes with the sparse access pattern introduces several new challenges in distributed training. Figure~\ref{fig:model_arch} shows a schematic of a deep learning based recommendation model. In a typical DLRM training setup, dense neural network (NN) parameters are replicated and stored on the GPUs and trained in data-parallel fashion, 
% and are replicated across trainers,   % (Figure~\ref{fig:existing_training_setup})
% The dense NN parameters are trained in a data-parallel fashion and 
where gradients are synchronized using the all-reduce communication collective. However, embedding tables are extremely large to hold in the GPU memory and are usually partitioned. 
% We next summarize three main approaches used by prior work to perform distributed training.

\noindent\textbf{Existing Systems.}
Several systems have been designed to perform offline recommendation model training due to it's popularity. 
% \noindent\textbf{Partitioned Embeddings.}
Training systems like TorchRec~\cite{torchrec}, FB-Research's DLRM~\cite{dlrmopensource} and HugeCTR~\cite{hugectr} partition the embedding table across different GPUs and train them in a model-parallel fashion. Embeddings are fetched using all-to-all collective~\cite{mpialltoall}.
While, TorchRec tries to overlap embedding-related operations, like remote embedding reads and writebacks, with the compute-intensive portion of the neural network, the amount of embedding data that needs to be fetched still adds significant overhead during training.
Figure~\ref{fig:time_distribution} shows a breakdown of the time taken for one iteration of training when using TorchRec~\cite{torchrec}. We observe that when using 8 training machines (AWS \emph{p3.2xlarge instances}), the overheads when compared to an ideal baseline that does not perform any embedding lookups, is around 70\% for the DLRM~\cite{naumov2019deep} and 75\% for DeepFM~\cite{guo2017deepfm}.

%TorchRec is another recently released industrial system  a system built over PyTorch, has recently been released. This system is heavily optimized for recommendation model training and . However, this overlapping does not reduce the magnitude of data fetched during training. 
Beyond spending a majority of time in embedding lookups, existing systems also couple storage and compute resources, \eg if the embedding tables for a model are extremely large but the compute requirements are  small, one still has to use a large number of GPU machines to store the embedding tables. This often leads to sub-optimal use of resources. 

% \noindent\textbf{Reordering Examples.} 
To alleviate the embedding access overhead and improve resource utilization, FAE~\cite{adnan2021high} performed an analysis of embedding accesses and observed
% shares a similar observation 
a similar skew in embedding access (\S\ref{sec:access_patterns}) patterns. However, FAE uses a reordering approach by dividing examples into hot and cold batches based on their embedding  accesses, this impacts the statistical efficiency as training continuously with hot batches changes the order of the training examples and can affect convergence~\cite{adnan2021high}. Further, it is not always possible to create batches that only access cached embeddings,  because some models~\cite{li2018learning, lee2018collaborative} use features like Unique User ID (UUID) or Session ID~\cite{tuan20173d, tan2016improved, twardowski2016modelling} that are unlikely to be repeated and thus requiring at least one cache miss per example.
Several other prior works~\cite{huang2021hierarchical,acun2021understanding, gupta2020architectural,miao2021het, lian2021persia} have proposed using asynchronous training to reduce embedding access overhead. With asynchronous training, embedding fetches can happen in the background, \eg in \het trainers can use embeddings that are stale up to a certain number of iterations. If embeddings are stale beyond the bound \het synchronizes those embeddings with the embedding server before a training iteration.
However, similar to other ML models, recent works~\cite{huang2021hierarchical,mudigere2021software} have observed that asynchronous training can lead to degradation in accuracy for recommendation models. Accuracy degradation is \emph{unacceptable} to large enterprises as it often directly leads to a loss in revenue~\cite{mudigere2021software}. 
% Further, in our experiments we find that the staleness bounds need to be tuned based on model architecture adding additional overheads for users (\S\ref{sec:existing_compare}).
Asynchronous training is also avoided due to the lack of reproducibility, which is necessary to reason about and compare different model versions. Therefore, in this work we focus on designing a system for \emph{synchronous distributed training}.
% \emph{Unlike prior systems, \system does not introduce any staleness and performs \emph{synchronous distributed training}}.
% \noindent\textbf{Compression Methods.}
% Other approaches improve performance using approximations such as tensor compression or gradient compression. Recent~\cite{yin2021ttrec} work reduces the size of embedding tables by almost 200x by using Tensor compression while using extra compute,
% while work by~\cite{gupta2021training} performs gradient compression to embedding updates to reduce synchronization time. Unlike \system{}, these approaches change the training algorithm and can lead to accuracy loss.

% Prior work~\cite{wilkening2021recssd, ke2020recnmp} has also used near data processing to try to reduce the overhead of embedding fetch by offloading some portion of the compute to the storage hardware DRAM and SSD. These optimizations are orthogonal to  \system.

%Another recent system \het~\cite{miao2021het} enables training of recommendation models with bounded asynchrony~\cite{li2014parameter};  Furthermore, to reduce embedding fetch overhead, \het caches most frequently used embeddings in a local cache present on the trainers.  Although, authors empirically show that such bounded asynchrony  does not lead to loss in accuracy, we find that the appropriate staleness bound depends on the model (\S\ref{sec:eval_converge}) and introduces additional tuning overhead for users. We focus 

% \paragraph{Training Setup.}
% There are two types

\begin{figure}[t]
    \centering
    % \begin{minipage}{0.48\textwidth}
    % \begin{center}
    % \includegraphics[width=0.7\linewidth]{images/current_setup.pdf}
    % \vspace{-0.1in}
    % \caption{\small{\textbf{Common distributed DLRM training setup}: The above figure represents a typical DLRM training setup. The dense NN layers are replicated across GPUs while the embedding layers are partitioned and kept in the main memory.}}
    % \label{fig:existing_training_setup}
    % \end{center}
    % \end{minipage}\quad
    % \begin{minipage}{0.48\textwidth}
    % \centering
    \includegraphics[width=0.6\linewidth]{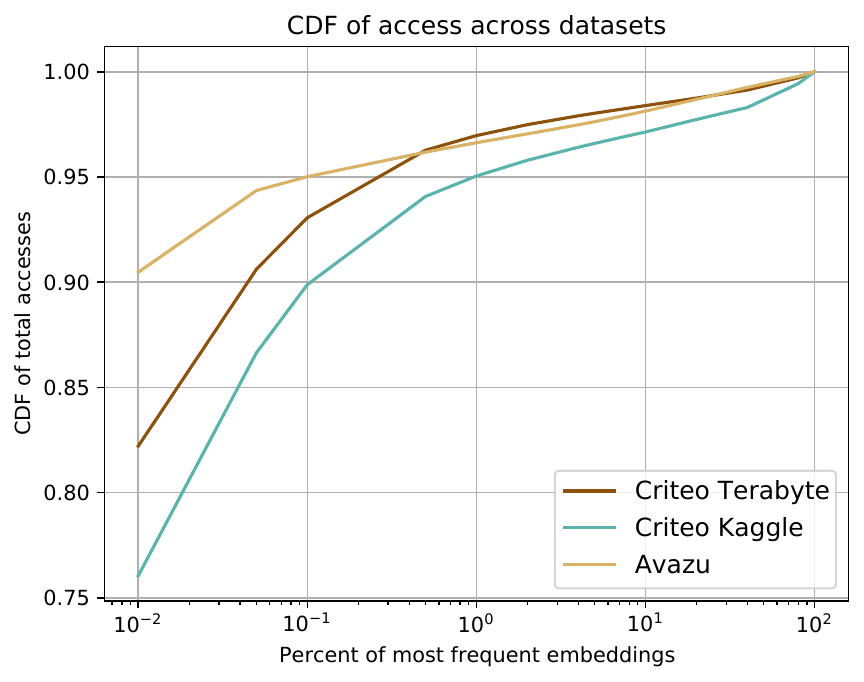}
    \vspace{-0.1in}
    \caption{\small{\textbf{CDF of embedding accesses:} The embedding access pattern is heavily skewed, with just the top 0.1\% of the embeddings responsible for more than 90\% of total accesses.}}
    \label{fig:access_cdf}
    % \end{minipage}
    \vspace{-15pt}
\end{figure}
 
    % \begin{figure*}[t]
    %     \includegraphics[width=0.9\linewidth]{hc_all.pdf}
    %     \centering
    %     \vspace{-0.1in}
    %     \caption{\small{\textbf{Popularity change across days}: We show the embedding popularity changes in the temporal domain. In the above figure we choose the popular embeddings on day zero and show how the number of accesses provided by the popular embeddings chosen on day 0 change over consecutive days. For some datasets we observed that the number of accesses provided by the cached popular embeddings decrease by as much as 20\%. This shows that we will have to compute our popular embeddings everyday if we want a high cache hit rate. }}
    %     \label{fig:popularity}
    %     \vspace{-10pt}
    % \end{figure*}

    \begin{figure*}[t]
    \includegraphics[width=0.8\linewidth]{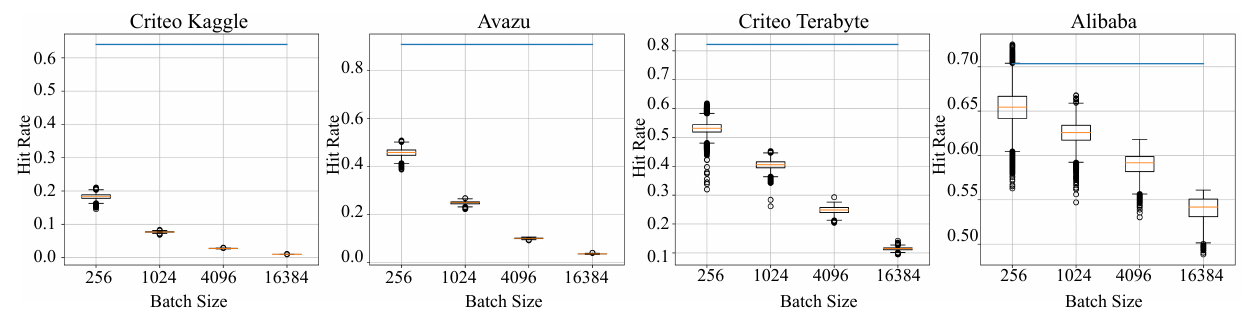}
    \centering
    \vspace{-20pt}
    \caption{\small{\textbf{Distribution of unique embeddings accessed from the cache}: 
    The hit rate of the cache decreases with increase in batch size.
    % The ratio of unique embeddings accessed from the cache decreases with increase in batch size. 
    The blue line represents the expected hit rate when storing the most frequently accessed 0.1\% embeddings. }}
    \label{fig:hit_rate_batch}
    \vspace{-10pt}
    \end{figure*}
%
%\vspace{-5pt}
\subsection{Embedding Access Patterns}
%\vspace{-5pt}
\label{sec:access_patterns}
Next, we describe some observations from analyzing embedding access patterns in recommendation model training.  

\noindent\textbf{Skew in embedding accesses.} When profiling embedding accesses across three datasets (details in Table~\ref{tab:datasets}), similar to prior work~\cite{adnan2021high}, we find that the embedding access pattern shows a large degree of skew. As shown in Figure~\ref{fig:access_cdf} we observe that almost 90\% of embedding accesses come from just 0.1\% of the embeddings.
% This is because, for example, in online shopping settings, some item IDs are accessed far more frequently because the corresponding items are more popular.
This indicates that caching \emph{hot} embeddings can reduce the number of embedding lookups significantly. Thus, when a training example needs a popular embedding, it can access it from a cache. 
% and \emph{cold} embeddings need to be fetched from the corresponding embedding server.  
However, we also observe that the set of popular embeddings can change over time. We measured the change in popularity over time across three datasets (Criteo, Alibaba, and Avazu) by, first choosing a fixed fraction of the most popular embeddings on the first day (\eg top 0.05\% most popular embeddings), and then measuring the percentage of accesses consisting of those embeddings in later days. 
We observed that such a scheme leads to a degrading cache hit ratio, 
% Using a fixed set of popular embeddings from the first day  would lead to a worse cache hit ratio on the following days
\eg in the case of the Avazu, if we chose the top 0.05\% of the most popular embeddings to be cached, the cache hit rate
% percentage of accesses to those embeddings
changes from 91\% on day 1 to 82\% on day 9, showing that static caching approaches~\cite{adnan2021high} will require
% it is not possible to use a fixed set of popular embeddings and existing approaches that focus on static caching~\cite{adnan2021high} will require 
regular updates for good performance.
The presence of similar skew in web scale data is  common~\cite{adamic2000power}. Prior work~\cite{leskovec2009community} also highlights that, at web scale, users create small communities and are more likely to strongly interact with members/items within those communities and very sparsely with items beyond their community. Therefore, we believe that most recommendation models trained on click stream data will have a similar skew as observed in Figure~\ref{fig:access_cdf}.

\paragraph{Long-tail of accesses limits benefits from caching.}
The analysis presented in previous paragraphs is for a scenario where embeddings are accessed for one example after another (i.e., batch size of 1). However, during recommendation model training it is common to use a large batch size (e.g., 16,384~\cite{mudigere2021software}). When embeddings are fetched for a batch, 
% we will only need to fetch the unique embeddings required for the batch. 
 fetching only the unique embeddings within a batch is sufficient.
Therefore, when calculating cache hit rates we should only account for the number of \emph{unique} accesses.

We next calculate the effectiveness of only caching the popular embeddings when using large batch sizes.
Figure~\ref{fig:hit_rate_batch} shows the distribution of \emph{unique embeddings} fetched from the cache as we increase the batch size. Most notably, we see that as batch size increases the ratio of embeddings fetched from the cache to the total number of unique embeddings (accounting for duplicates across examples in a batch) needed in a batch keeps decreasing drastically, \eg for a batch size of 16,384 (standard batch size used in MLPerf~\cite{mattson2020mlperf}), only around 10\% of the total unique embeddings required are fetched from the cache. 
% As the batch size increases, we find that a large number of embeddings are still required to be fetched from outside the cache thereby affecting training throughput.
 Quantitatively, for the Criteo dataset, we observed that a batch of 16,384 needs 425,984 embeddings. Within this batch there are only around 65,000 unique embeddings (or unique categorical features); out of these 65,000 embeddings, the cache only has a 10\% hit rate.

\subsection{Design Goals}
% \vspace{-5pt}
Based on the above discussion we have three primary design goals: (i) High throughput training of deep learning based recommendation models by reducing the embedding access overhead, (ii) Provide the same guarantees as synchronous training to improve reproducibility and maintain training accuracy, and (iii) Design a disaggregated architecture where memory and compute can be independently scaled. 

We approach our design based on the workload patterns observed and aim to improve performance by speeding up access to both frequently used and long-tail of embeddings. We design \system,  a framework that caches hot embeddings and performs out-of-order prefetching for the long-tail embeddings. We describe our design next. 
\section{\system Design}
% \vspace{-3pt}
%\saurabh{note for continuity: Change the previous paragraph to flow to finish with saying caching alone with not suffice, we need a system which can handle both hot accesses and long tail of accesses and that system will need to use both caching an prefetching, caching is justified for prefetching say that long tail access embedding, discuss system goals - reduced embedding access overhead, synchronous training, disaggregated architecture.}

% We begin by presenting our design goals for \system: 

% \shivaram{this can go into end of motivation as 2.3 design goals.}

% As discussed in the previous section, we need a system which can support both caching and prefetching.

%\system is a large scale recommendation model training system  %note I want to explicitly say somewhere that there is a cache 
%which speeds up training by reducing the overhead of embedding accesses. 
%Unlike prior systems~\cite{adnan2021high, miao2021het} which only speed up access of frequently used (hot) embeddings, \system also speeds up accesses of long tail embeddings by out of order prefetching, while providing the same guarantees as synchronous training.  
%To speedup access of long tail embeddings, %\system relies on the insight - \emph{if an element is in the long tail it would have not been updated recently}. 
We begin by providing an overview of our design.
% In this section, we first an overview of our system design and then describe mechanisms using which \system is able to significantly reduce the overhead of embedding accesses and provide high throughput recommendation model training. 
% Next we present the design of \system, a system that can accelerate training of large-scale recommendation models.
% \vspace{-5pt}
\subsection{Design Overview}
% \vspace{-5pt}
% \system first we introduce the idea \emph{lookahead} where based on the 

\begin{figure}[t]
    \centering
    \includegraphics[width=\linewidth]{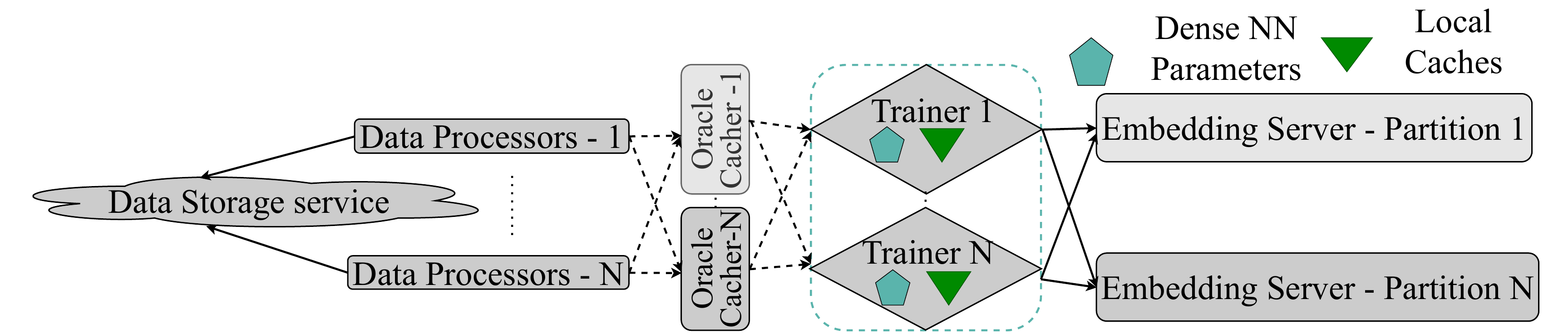}
    \vspace{-25pt}
    \caption{\small{\textbf{\system{} setup:} All the components of \system can be individually scaled. The dashed arrows signify async RPCs while solid ones signify sync RPCs.}}
    \label{fig:bagpipe_training_setup}
    % \vspace{-0.2in}
    \vspace{-10pt}
\end{figure}

%Before discussing our contributions in detail, we briefly introduce the reader to the high level design of \system. As shown in 
% is a component which is specific to \system, it is where \system 
%It is very important that \orc is extremely low overhead to scale \system. 
% At a high-level,
\system consists of four components that collectively perform training as shown in Figure~\ref{fig:bagpipe_training_setup}. Each iteration of training begins with sampling a batch of examples, the \texttt{DataProcessors} pre-process the examples and send them to the \orc. Based on the examples, \orc runs a  \emph{lookahead algorithm} to determine embeddings to prefetch and to cache, and dispatches this information to the \texttt{Trainers}. The trainers typically run on GPU machines and perform gradient computation. Trainers hold the dense parameters of the model and \system's cache in the GPU memory. Also, trainers fetch necessary embeddings from the \texttt{EmbeddingServers}. Embedding servers hold the embedding tables of the recommendation model.
This design introduces the following contributions:
%We discuss the design and it's benefits in more detail in \S\ref{sec:system_design} 

% To decide what to cache and prefetch
%cache design. Specifically, we introduce
\begin{myitemizeleft}
\vspace{-3pt}
     \item \system utilizes both caching and prefetching to reduce embedding access overhead. Given an offline training regime, we introduce the concept of \emph{lookahead}, where we can look beyond the current batch and decide which elements to cache and prefetch (\S\ref{sec:cache_prefetch_bp}).
     % , given offline training
     % despite performing caching and out of order pre-fetching
     % % we introduce a partially synchronized \emph{oracular} cache. To reduce the overhead of synchronization of caches while minimizing the control logic
    \item We extend our scheme to the distributed setting and introduce a logically replicated, physically partitioned cache design (\S\ref{sec:distributed_cache}) to minimize communication overheads. 
    % % that can retain the same semantics as synchronous training.
     To further reduce synchronization overheads we use CPA,
    % \footnote{based on the just-in-time inventory management system pioneered by Toyota.} 
    to selectively synchronize parts of the cache that are immediately needed on the critical path while synchronizing the rest  in the background. %(\S\ref{sec:bagpipe_overlap}). 

    \item Finally, we discuss how \system's dis-aggregated design can help improve efficiency, by scaling components depending on the properties of the dataset and the model, and enable low-overhead fault tolerance (\S\ref{sec:ft}).
\end{myitemizeleft}
%only the part which
% \saurabh{I have a flow problem here. some part of cache needs some sort of design intro and I am stuck with this problem for more than half a day. }

% which is 
% to support larger embedding tables and trainers 
% independently 

 % \begin{figure}

 % \begin{minipa}
 % \begin{minipage}[t]{0.9\textwidth}
\begin{algorithm}[t]
% \scriptsize{}
\small
\caption{Lookahead Algorithm}
\label{alg:lookaheadalg}
\SetKwInOut{KwIn}{Input}
\KwIn{LookAheadValue}

$BatchQueue = Queue()$\;
$LatestTracker = Dictionary()$\;
$InCache = Set()$\;

\While{$\mli{BatchQueue}.size() > 0$} {
 \While{$hasNextBatch()$ and $BatchQueue.size() < LookAheadValue$}{
         $Batch = getNextBatch()$\;
        $IterationNumber = Batch.IterationNum$\;
        \For{\texttt{$\mli{EmbID} \in  Batch.UniqueEmbeddings()$}}{
            $LatestTracker [\mli{EmbID}] = IterationNumber$ \;
            $BatchQueue.pushBack(Batch)$\;
        }

 }
 $\mli{TTLUpdateRequests} = \{\}$\;
        $CacheFetchRequests = \{\}$\;
        $CurrentBatch =BatchQueue.popFront()$\;
        \For{\texttt{$\mli{EmbID} \in  CurrentBatch.UniqueEmbeddings()$}}{
            $\mli{TTL} = LatestTracker [\mli{EmbID}]$\;
            $\mli{TTLUpdateRequests}.append((\mli{EmbID}, \mli{TTL}))$\;
                    \If {$\mli{EmbID}$ \textnormal{is not in} $InCache$} {

                $CacheFetchRequests.append(\mli{EmbID})$\;
                $InCache.insert(\mli{EmbID})$\;
        }
        \If {$\mli{TTL} = CurrentBatch.number$} {
                $InCache.erase(\mli{EmbID})$\;
                $LatestTracker.erase(\mli{EmbID})$\;
        
        }

        }
                $SendToTrainers(\mli{TTLUpdateRequests})$\;
        $SendToTrainers(CacheFetchRequests)$\;
}
% \v
\end{algorithm}
% \end{minipage}
% \vspace{-10in}
% \end{figure}
% one of the hardest problems is to
% \vspace{-5pt}
\subsection{Caching and Prefetching in \system}
% \vspace{-5pt}
\label{sec:cache_prefetch_bp}
In \system, we introduce the idea of each trainer having a \emph{local} cache. When designing a system with caching, we need to design a cache insertion policy (what to cache?) and a cache eviction policy (what to evict?) so as to maximize the hit rate. However, offline batch training of machine learning jobs like recommendation model training has additional structure: \emph{in offline batch training future batches and their contents are predictable}, \ie in context of recommendation models we can look beyond the current batch and infer which embeddings will be accessed in future batches. This insight helps us create a perfect or \emph{oracular cache}. 
To utilize this insight, we design a lookahead algorithm. 
% We utilize this insight by designing a lookahead algorithm which we discuss next. 
For ease of explanation, the discussion in this section assumes there is only one trainer (only one cache). We extend this to the distributed setting in \S\ref{sec:distributed_cache}.

\noindent\textbf{Lookahead Algorithm.}
To decide what to cache and what to evict we develop a low overhead (benchmark in \S\ref{sec:oracle_cacher_benchmark}) lookahead algorithm (Algorithm~\ref{alg:lookaheadalg}) which also ensures consistent access to embeddings. We denote the \emph{lookahead value} (\lkval), as the number of batches 
% ahead of 
beyond the current batch, 
% we are going 
which will be analyzed to determine what to cache, \eg if the current batch is $x$, we consider embedding accesses in batches from $x$ to $x$+\lkval, to determine which elements in batch $x$ should be cached.
The lookahead algorithm takes three inputs: the current batch, future batches  (next \lkval number of batches), and current state of the cache on the trainer. 
% To apply thi, we create a low overhead lookahead algorithm  (Algorithm~\ref{alg:lookaheadalg} which can \emph{lookahead} beyond current batch and decide which elements to cache and which to prefetch based on future access patterns.
% Specifically we denote \lkval, or lookahead value, as the number of batches ahead of the current batch we are going to look at.
% Given this approach we next discuss how a cache can be maintained in a single trainer setting (distributed setting discussed in \S\ref{sec:distributed_cache}). 

The lookahead algorithm outputs two pieces of information. %\shivaram{make it two pieces and say everything is just ttl?}
First, for the current batch, it generates the list of embeddings that will not be found in the cache on the trainer's GPUs. This allows \system to \emph{prefetch} these embeddings out of order before the current batch is used for training. Prefetching allows \system to hide the data access latency for the long tail of embeddings that are not frequently accessed. 
% early
%This guarantees that when batch is loaded for training it will find all the necessary embeddings in the dynamic local cache. 
% within the future batches
Second, the lookahead algorithm determines \emph{which embeddings \textbf{from the current batch} will be used in future batches}, and the last iteration they will be accessed in the current lookahead window. Any embeddings from the current batch that will be used by future batches in the \emph{lookahead} window will be marked for caching, so they can be accessed from the GPU memory in the future. The last iteration an embedding is used within the \emph{lookahead} range and serves as time-to-live (TTL) for the embedding in the cache.

% Our lookahead algorithm (Algorithm~\ref{alg:lookaheadalg}) dynamically keeps a window of batches of that size. The window is maintained as a queue, where the next batch to be processed is in the front of the queue. As shown in Line 5-12, when the window size is smaller than the look-ahead value (at the beginning of the training or after a batch is processed), the algorithm extends the window by reading new batches from the data loader. Meanwhile, it also keeps the last occurrence of each embedding in this window. After ensuring that the window size is equal to \lkval (or there are no more batches to process), the algorithm starts to process the first batch in the window, as shown in Line 15-27. For each embedding in the batch, we add a TTL update to denote the latest occurrence of this embedding, and then add a pre-fetch request if this embedding is not in cache.

 \begin{figure*}[t]
    \includegraphics[width=0.9\linewidth]{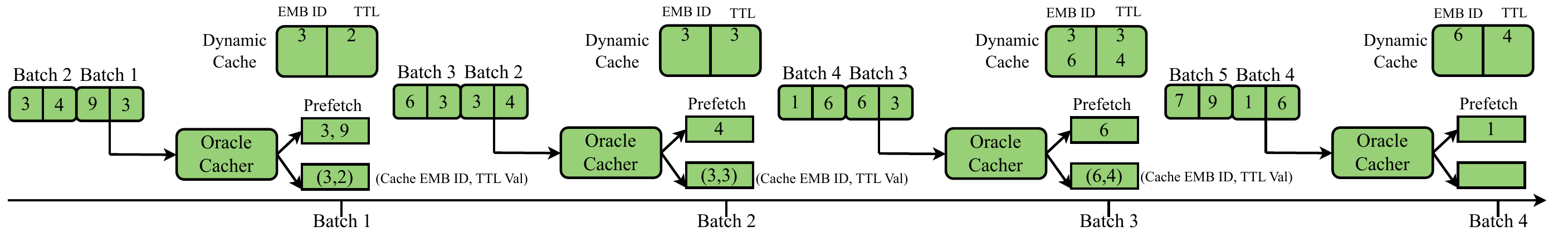}
    \centering
    \vspace{-0.12in}
    \caption{\small{\textbf{Lookahead Algorithm}: The above figure shows an illustration at different batch steps of how the lookahead algorithm functions. In the above example, the lookahead value is 2 and the batch size is also 2.}}
    \label{fig:oracle_cacher_example}
    \vspace{-10pt}
    \end{figure*}
\noindent Next we describe an example of how lookahead algorithm processes batches (Figure~\ref{fig:oracle_cacher_example}), with lookahead value (\lkval) as 2: %and show how a window of batches is processed by the lookahead algorithm. 
\begin{myitemizeleft}
\vspace{-10pt}
  % \item \textbf{Batch 1} Embedding 3 and 9 are in the batch. For both embedding 3 and 9, we will launch prefetches. However, for embedding 3, we see that it is accessed again, and the last occurrence in the window is at Batch 2, so we decide to cache it with the TTL set to 2. 
  \item \textbf{Batch 1} Embedding 3 and 9 are in the batch. For both embeddings we launch prefetches. However, embedding 3, is accessed again, and the last occurrence in the window is at Batch 2, so we cache it with the TTL set to 2.
  
  % \item \textbf{Batch 2} For embedding 3, we see that it is in the cache but the last occurrence in the window is in now Batch 3. So we will send a TTL update for 3. We will not launch a prefetch for 3 because 3 is already in the cache, while we will prefetch 4 since it is not in the cache. 

  \item \textbf{Batch 2} Embedding 3 is in the cache so we do not send a prefetch request. But the last occurrence for 3 in the window is now in Batch 3. Therefore, a TTL update is sent for 3. We will prefetch 4 since it is not in the cache.

  % \item \textbf{Batch 3} For embedding 3, there are no future occurrence in the window, so we will not send any TTL updates. For embedding 6, we decide to cache it with a TTL of 4 because it will be used in Batch 4. Embedding 3 will be evicted from the cache after this batch.
    \item \textbf{Batch 3} We prefetch embedding 6 and cache it with a TTL of 4 since it will be reused in Batch 4. At this point in our lookahead window Embedding 3 has no future occurrence so it will be evicted after batch 3. However, if embedding 3 was being used by batch 4, we would have kept it in the cache and sent a TTL update with eviction batch as 4.

        % \item \textbf{Batch 3} Embedding 6, will be prefeteched and laa prefetch and cache it with a TTL of 4 since it will be reused in Batch 4. Embedding 3 has no future occurrence so it will be evicted after batch 3.
    % For embedding 3, there are no future occurrence in the window, so we will not send any TTL updates. 
  
  % \item \textbf{Batch 4} For this iteration, we will prefetch 1 since it is not in the cache. We do not send any TTL updates for 6 as it is absent in future batches. It will be evicted from the cache after this batch.

\item \textbf{Batch 4} We prefetch 1 since it is not in the cache. We do not send any TTL updates for 6 as it is absent in future batches and will be evicted  after this batch.
  \vspace{-5pt}
  \end{myitemizeleft}

\noindent\textbf{Consistency with the Lookahead algorithm.}
Our consistency goal is to avoid staleness and ensure that trainers do not prefetch an embedding from the embedding servers while it has updates that have not yet been written back.
Despite pre-fetching embeddings out of order, our formulation of what to cache and what to prefetch (Algorithm~\ref{alg:lookaheadalg}), provides an extremely important guarantee that allows us to maintain consistency and match the execution of synchronous training. 
 % minimize communication overheads whil
 % lookahead algorithm
 When the trainer is processing batch number $x$, an embedding used by the batch will either be available in the cache with it's most recent value or no preceding batch in the lookahead range (any batch number in [$x - \text{\lkval}, x)$) would have updated that specific embedding. That is, if an embedding was needed by a batch in batch number in range $[x-\text{\lkval}, x)$ it will be in the cache; if an embedding is not in the cache it means no batch in range of $[x-\text{\lkval}, x)$ has updated it. Therefore, as long as the prefetch request for batch $x$ is issued after updates from training batch number $x - \text{\lkval}$ have been written back, we can guarantee that we will not see stale embeddings.
 %and hence obtain the same results as synchronous training. 

%  This condition guarantees that despite fetching embeddings out of order
% can safely prefetch embeddings without being concerned about staleness
 
% Guarantees provided by caching
% to distributed setting
% \vspace{-5pt}
\subsection{Distributed Cache Design in \system}
% \vspace{-5pt}
First we discuss the requirements for the distributed cache design and our goals. Next, we discuss the design space for cache design and finally we compare these designs both quantitatively and qualitatively.
\label{sec:distributed_cache}

\noindent\textbf{Distributed Cache Requirements.}
When extending the caching scheme described above to a distributed setting, \system can provide consistency as long as the following two requirements are satisfied
% In order to extend the caching scheme described above to distributed settings while still guaranteeing consistency. We observe that in a distributed setting, \system can provide 
% consistency as long as the following two requirements are satisfied: 
(i) Each trainer sends prefetch requests for batch number $x$ only when cache eviction and updates have been performed by \emph{all} the trainers on $x-\text{\lkval}$ batch. (ii) Each trainer's cache should contain the latest value of the embedding.
The first requirement is a direct extension of our prior discussion and can be satisfied by synchronizing the iteration number that each trainer has processed. The second condition, however, creates additional communication overheads and we next discuss the design space and techniques to reduce these overheads. 

% toward the end of previous sub-section (\S\ref{sec:cache_prefetch_bp}). We can easily satisfy it,
% configuring the trainers to launch prefetch based on this required condition
% . We can satisfy this easily as trainers need to synchronize their dense ML parameters in each iteration and 

% and needs to be designed much more carefully so as to minimize this communication overhead. In next several paragraphs 
% in \system
%of synchronization.
% guaranteeing consistency
%challenges in distributed setting. 

%\paragraph{Cache Design}
%\saurabh{need to make this a subsection somehow}
\noindent\textbf{Goals of distributed cache design.}
The primary objective of our distributed cache design is to minimize the \emph{time spent} on cache synchronization on the \emph{critical path}. Thus our objective includes, accounting for the number of bytes transferred (bandwidth) and connection overheads (latency)~\cite{thakur2005optimization}.  
% More than just minimizing the bytes communicated during synchronization our design should aim to minimize the total wall-clock time spent during synchronization. 

Next, we explore the distributed cache design space and discuss synchronization costs with each design. 
% main design choices; a replicated cache, a partitioned cache, or a hybrid variant in between replicated and partitioned. \sosp{Next we introduce these design choices and post that we discuss their communication and im}

%In next several paragraphs we will discuss various options for performing distributed caching in \system.
% In case of replicated cache, the control logic is significantly simplified as each element is available locally on all the machines so there is no need to keep a record of the location of each embedding.

\noindent\textbf{Replicated Cache.}
In a replicated cache, each trainer will pre-fetch \emph{all} the embeddings which are required by the whole batch (not just a worker's partition of the batch). After performing the backward pass we synchronize all the elements which have updated gradients across all the workers, such that embeddings in the caches are synchronized at the end of each iteration using \emph{all-reduce}. This trivially ensures that 
each trainer's cache has the latest version of the embedding.
%In this setup, the trainers do not need to perform communication for the forward pass on the critical training path. 
A replicated design results in high bandwidth cost due to synchronization of all the elements across all trainers even if the element's updated value would not be required in future by other trainers, \ie it will be evicted from the cache. However, there is very small control (latency) overhead because all elements are synchronized.
% We observe that for an 8-machine setup with Kaggle Criteo Dataset around 65.6K embeddings per machine need to be synchronized using all-reduce primitive. 
%Further, the synchronization in a replicated cache can be performed with efficient \emph{all-reduce} primitives. 
% Using the equation for amount of data transferred per node in case of all-reduce~\cite{thakur2005optimization} we observe this to be around 20 MB.  
% Qualitatively each of these approach has different benefits. 

% Before comparing partitioned cache with replicated cache, we would like to discuss additional optimizations which can be performed in both cases to reduce the overhead of cache synchronization. 
%options for cache design
%choosing different options
% evaluating different options
% Implementing different options

%  of a replicated cache. In case of a partitioned cache 
% from which it fetched the embedding 
\noindent\textbf{Partitioned Cache.}
A partitioned cache is on the other end of the design spectrum where each trainer is assigned an \emph{exclusive} portion of the cache. Before the forward pass of training, each trainer fetches the embeddings not available locally from their peer trainers. Post backward pass, each trainer writes back the gradients to the respective peer trainer which has ownership of the embedding. These steps are required to ensure we always use the latest version of the embedding.
Unlike the replicated cache, where all the embeddings are synchronized irrespective of whether a trainer needs it, in the case of a partitioned cache, trainers only fetch and write back the embeddings they utilize. 
Further, partitioned caches are more space efficient as there is only one copy of each embedding in the distributed cache.

% \noindent\textbf{Partitioned Cache-Communication Aware.}
The number of bytes communicated when using a partitioned cache depends on how batches are partitioned across trainers. 
To study the scenario where batch partitioning is communication aware, \ie batches are partitioned so as to minimize bytes communicated across trainers, 
% For Kaggle Criteo dataset with  8 trainers, we observed approximately 24940 embeddings per machine need to be synchronized using p2p connections.
%However, we observe in Figure~\ref{fig:all_lkval_effect} the cache space requirement is quite small compared to memory available and we are more concerned with the cache synchronization overhead. 
% \noindent\textbf{Partitioned Cache with Optimal Batching.}
%In case of a partitioned caches the amount of communication is dependent on batch partitioning scheme, \eg if the partitioning scheme assigns batches so as to minimize inter-node embedding accesses it can potentially reduce the inter-node embedding accesses. 
% To derive an optimal partitioning scheme 
we formulate a mixed integer linear program (MILP). Given the cache state on all trainers, the MILP computes a partitioning of examples which minimizes the amount of inter-node communication.  
Given a batch of examples ($b$) and $p$ trainers, we introduce $b\times p$ variables in our MILP. Each variable is denoted by $x_{i,j}$ where if $x_{i,j} = 1$, then example $i$ will be assigned to trainer $j$. We then compute a cost matrix $C$, where given the cache state, $C_{i,j}$ represents the cost of inter-node communication that will be required to fetch embeddings for example $i$ to location $j$. 
Our objective is to minimize the amount of inter-node communication. We formulate it using our variables and cost matrix as: 
% \setlength{\abovedisplayskip}{1pt}
% \setlength{\belowdisplayskip}{1pt}
% \setlength{\abovedisplayshortskip}{1pt}
% \setlength{\belowdisplayshortskip}{1pt}
% \begin{equation}
    $$\text{Minimize} \quad \sum_{i \in I} \sum_{j \in J} C_{i,j} \cdot x_{i,j} $$
    % \end{equation}
Where $I$ and $J$ represent the set of examples and trainers respectively.
Further, we include two constraints to ensure that the solution is feasible and avoids load imbalance:
% \begin{myitemize}
%     \item Each example must be placed on one trainer and all examples need to be placed on at least one trainer node.
%     \begin{equation}
%        \forall i \in I \quad \sum_{j \in J} x_{i,j} = 1
%     \end{equation}
%     \item Next we want to make sure that the batch is equally distributed across machines. For this, we include a constraint that the number of examples is evenly partitioned. 
%     \begin{equation}
%     \forall j \in J \quad \sum_{i\in I} x_{i,j} =  \forall \hat{j} \in J \backslash \set{j} \quad \sum_{i \in I} x_{i,\hat{j}}
%     \end{equation}
% \end{myitemize}

% \begin{myitemize}
(i) Each example must be placed on one trainer and all examples need to be placed on at least one trainer node.
    % \begin{equation}
     $\forall i \in I \quad \sum_{j \in J} x_{i,j} = 1.$
    % \end{equation}
(ii)  We add another constraint to make sure the batch is equally distributed across machines to prevent load imbalance. 
% For this, we include a constraint that the number of examples is evenly partitioned, 
    % $\forall j \in J \quad \sum_{i\in I} x_{i,j} =  \forall \hat{j} \in J \backslash \set{j} \quad \sum_{i \in I} x_{i,\hat{j}}$ 
% \end{myitemize}
The optimization problem can be solved using existing MILP solvers like Gurobi~\cite{gurobi}.
% and is effective at reducing the number of embeddings synchronized. 
% For the Kaggle Criteo Dataset with 8 trainers, we find that for communication aware paritioned scheme each machine needs to synchronize around 21K embeddings when compared to 65.6K embeddings with a replicated cache. 

However, using communication aware partitioned caches has two disadvantages: first solving the MILP takes around 2.36s on a 16-core machine making it infeasible when iteration times are around 100ms (Figure~\ref{fig:time_distribution}).  Secondly, sync time does not solely depend on bytes communicated, as overheads from maintaining data-structures and establishing connections also play a role. With partitioned caches we would need to introduce additional data structures to keep track of embedding locations and establish multiple connections.
% need to be fetched from which peer.
% as the \orc performs partitioning before examples are dispatched to the trainers, sending the partition mapping increases the communication overhead between the \orc and trainers
%is when using a partitioned cache, 

% \begin{figure*}[t]
%     \includegraphics[width=0.7\linewidth]{images/forward_pass_schematic_no_white.pdf}
%     \centering
%     \vspace{-0.1in}
%     \caption{\small{\textbf{Schematic delayed synchronization}: The above figure shows an illustration of how \system manages to minimize the communication and cache management on the critical path. \system only synchronizes those embeddings  on the critical path which are going to be used by the next iteration. In iteration 1 we only synchronize embeddings needed in iteration 2 ($E_{iter_2}$). Rest of the embeddings ($E\notin E_{iter_2}$) are synchronized off the critical path. Before every synchronization on the critical path, we have a barrier to ensure prior synchronizations have finished. 
%     }}
%     \label{fig:overlap_schematic}
%     \vspace{-15pt}
%     \end{figure*}

\begin{figure}[t]
\centering
\begin{minipage}[t]{0.47\linewidth}
\includegraphics[width=1.1\linewidth]{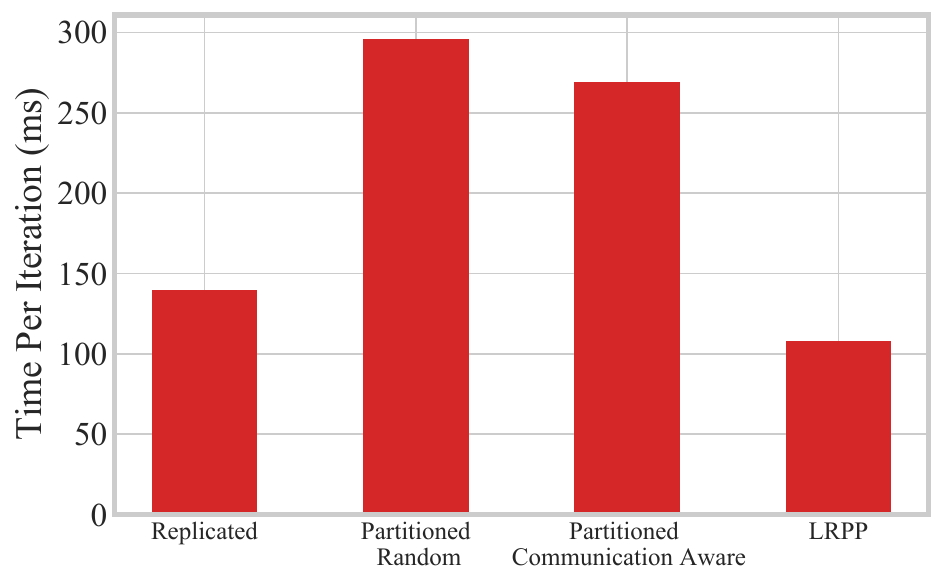}
\vspace{-15pt}
    \caption{\small{\textbf{Comparing cache designs:}We observe that LRPP provides best performance among all other cache options.}}
    \label{fig:effect_per_iter_time}
\end{minipage}\quad
\begin{minipage}[t]{0.47\linewidth}
\includegraphics[width=1.1\linewidth]{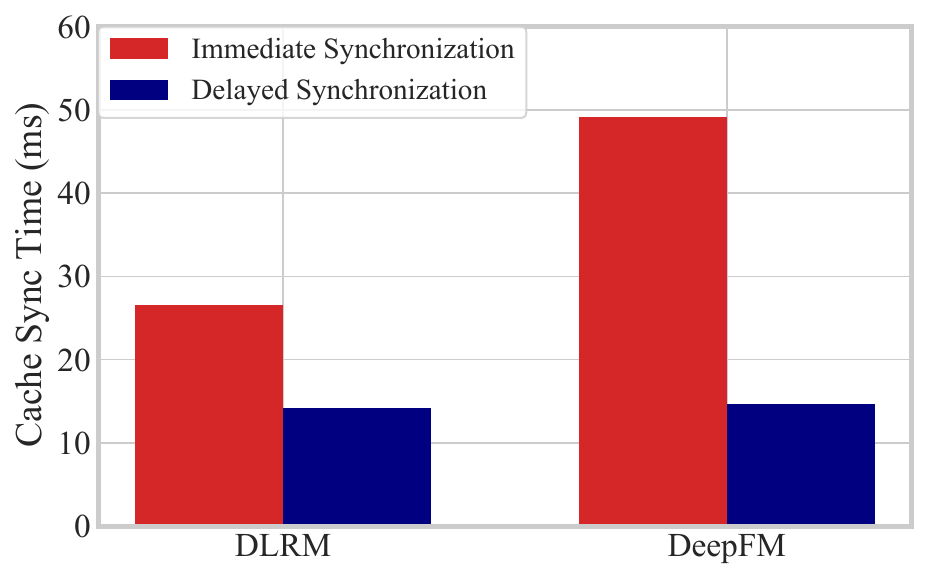}
 \vspace{-15pt}
     \caption{\small{\textbf{Effect of Delayed Synchronization}:}Delayed Sync can reduce time for cache synchronization by up to 44\%. }
    \label{fig:split_diff}
\end{minipage}
% \quad
% \begin{minipage}[t]{0.32\textwidth}
%  % \includegraphics[width=0.9\linewidth]{images/compare_hardwares.pdf}
%     \includegraphics[width=0.9\linewidth]{images/benefit_kandban_lrpp.pdf}
%     \vspace{-0.1in}
%     \caption{\small{\textbf{Benefits of Cache Optimizations:} Use of LRPP and delayed synchronization can reduce the synchronization on critical training path by up to 70.3\%.}}
%         \label{fig:cache_benefit_overlap}\quad
% \end{minipage}
\vspace{-15pt}
\end{figure}

\noindent\textbf{Logically Replicate Physically Partitioned Cache.}
%Each of the above caches have their drawbacks, replicated caches perform synchronization of embeddings which might not be useful thus have high synchronization overhead. While, partitioned caches need additional control logic (which slows down \orc) and still need to communicated embeddings to their peer. 
Ideally, we would like to design a cache that does not perform unnecessary synchronization of embeddings but does not introduce additional overheads due to state tracking. 
%Further, in \system we are more concerned about cache synchronization phase as we can easily overlap the pre-fetch portion. 
To achieve this goal, we propose using Logically Replicated, Physically Partitioned (LRPP) caches. By \emph{logically replicated} we mean that from the view of \orc all caches have all data and are fully replicated but by being \emph{physically partitioned}, the trainers decide which elements need synchronization and which elements can be evicted without synchronization. The primary insight behind our idea comes from the observation that for the Criteo dataset, around 25\% of the embeddings are used by only one of the examples in batch. Therefore, these embeddings are updated at only one trainer before being evicted. Thus, fetching or synchronizing them across all trainers is a waste of network bandwidth. We design a new protocol that modifies the replicated cache based on this insight. 

With LRPP caches, the \orc marks embeddings which are only used by a single trainer. Given this metadata, these embeddings are only fetched by the trainer which needs them and are ignored by other trainers. After the forward and backward pass completes, the trainers skip synchronization for these embeddings and use all-reduce to synchronize the other embeddings. In the background, the trainer which made the only update to the marked embedding evicts it back to the \embs. This optimization is able to reduce the volume of embeddings prefetched and synchronize with very minimal control logic. LRPP can be further extended with more fine-grained partitioning, \ie we can synchronize embeddings updated by two workers using a separate communication group containing just those workers. However,  further fine-grained partitioning will create additional control logic, which in turn would add additional latency and thus yield diminishing returns.
% further 
%we will observe diminishing returns. 
% Further, with just LRPP we were able to reduce embedding access overhead down to around 10\% 

There are parallels between design of  LRPP  to ~\cite{yang2023fifo} a concurrent work which  only caches elements which are going to be utilized more than once with a FIFO eviction policy. This is analogous to LRPP only synchronizing elements which are  going to be used in future.
We plan to study extensions of LRPP in future work.

\noindent\textbf{Comparing cache design choices.}
For Kaggle critieo dataset with batch size 16,384 and 8 trainer machines (p3.2xlarge) we observe that Replicated Cache communicates around 65K embeddings per iteration, while Communication Aware-Partitioned Cache communicates 21K embeddings per iteration and LRPP communicates around  48K embeddings. 
 % For the same Kaggle Criteo dataset, with LRPP, around 48K embeddings need to be synchronized with 8 trainers. 
Further, we implement all these  in \system and evaluate them in terms of per-iteration training time using the same setup. To consider the best case scenario for partitioned caches, we ignore the time taken by the Gurboi solver.
In Figure~\ref{fig:effect_per_iter_time}, we observe that LRPP outperforms replicated by 22.8\% and communication aware partitioned by 59.8\%. Our analysis shows that despite synchronizing fewer embeddings partitioned caches do not perform well due to hotspots and additional control logic. Since some embeddings are accessed extremely frequently, the trainers that own those embeddings become a bottleneck. Further, in partitioned caches, the overhead of performing multiple collective communication calls, creating memory buffers for each collective communication call and tracking which peer to access embeddings from, leads to an additional overhead of 80-90ms for a batch size of 16K with 8 trainers.
Therefore, we configure \system to use LRPP cache synchronization scheme due to it's superior performance.

% \section{}
\noindent\textbf{Delayed Synchronization.}
To further optimize the LRPP protocol we use Critical Path Analysis~\cite{yang1988critical}.  In this scenario, CPA implies that as long as the embeddings are synchronized before being critically required it can suffice. However, directly using CPA in context of embedding synchronizations for recommendation models will lead to network contention with other competing synchronizations. Therefore, we introduce delayed synchronization, where we only synchronize the embeddings which will be required in the next iteration on the critical path. The embeddings which are not needed immediately are synchronized in the background. To, avoid network contention due to background synchronization we ensure that
all background synchronizations are completed before we launch other critical path synchronizations for future iterations. 
On Kaggle Criteo dataset with 8 trainers and batch size 16K, we see that only 22.7K embeddings out of 48K embeddings (47.3\%) need to be synchronized on the critical path, the rest can be overlapped with the forward pass of the next iteration. For two models on Criteo dataset, Figure~\ref{fig:split_diff} shows that delayed synchronization can further reduce cache synchronization time by up to 44\% (in addition to LRPP) by overlapping synchronization with forward pass.
We also observe that LRPP and delayed synchornization can together reduce bytes communicated on critical path by around 70\%.

\subsection{Disaggregated Design and Fault Tolerance}
% \vspace{-5pt}
\label{sec:system_design}
\label{sec:ft}
Existing recommendation model training systems ~\cite{dlrmopensource, mudigere2021software, torchrec} couple storage and compute resources, \ie it is not possible to scale the number of embedding table partitions without increasing the number of trainers.  This affects fault tolerance and resource utilization. For fault tolerance, given the extremely large embedding table sizes, checkpointing a trainer can take several minutes~\cite{eisenman2022check} during which the compute resources stay idle.
The lack of \emph{disaggregation} also leads to poor resource utilization~\cite{dageville2016snowflake, armbrust2021lakehouse}, \eg when embedding tables are extremely large but the dense neural network parameters are small, an optimal configuration would be to use more servers for embedding tables but have fewer trainers.
Thus, we design a disaggregated architecture for \system
(Figure~\ref{fig:bagpipe_training_setup}) with four major components: (i) Data Processors (ii) Oracle Cacher (iii) Distributed Trainer (iv) Embedding Servers.

\noindent\textbf{Data Processor.} Data processors read  and batch training data 
which is resource intensive. 
Similar to prior designs~\cite{zhao2021understanding} we offload data processing to reduce  trainer overheads. Data processors are stateless and can be restarted on failure.

\noindent\textbf{\orc.}
\orc is a centralized service that inspects all the training batches  using the lookahead algorithm (Algorithm~\ref{alg:lookaheadalg}).  \orc decides which elements to prefetch for the current batch and the TTL for eviction of elements being cached.
\orc sends the training data as well as the embedding ids that need to be cached/prefetched using async RPC calls to the trainers. \orc is designed such that all the necessary internal state is also present on the trainers. Therefore, whenever \orc has to be restarted we only need to fetch the last iteration number processed by the trainers and the embedding IDs present on them. 
% prefetch requests,
%An important point to note is that the cache changes for an iteration do not happen when the oracle cacher generates the requests. Those requests are sent to the trainers with the iteration number to apply them at. The requests are then applied by the trainer at an appropriate time so as to maintain consistency. 

% Trainers also hold the LRPP cache in GPU memory where the embeddings needed by future batches are stored. 
\noindent\textbf{Trainer.} Trainers hold the dense neural network portion of the recommendation model and the LRPP cache in the GPU memory. The trainers perform forward and backward passes in a synchronous fashion. Trainers also: (i) prefetch the embeddings based on requests sent by \orc (ii) perform cache maintenance including addition and eviction of embeddings. When a trainer fails, \orc makes an RPC call to ask existing trainers to checkpoint their state (model parameters and cache contents) and then copies this state to the newly started trainer. Each of the trainers then discard their gradients and \orc starts from the previous iteration.  With delayed synchronization and LRPP enabled we might loose updates of at most one iteration, which is unlikely to affect model convergence~\cite{stich2018local}.

% of the model
% We evaluate the overhead of this in \S\ref{sec:eval}.

%and updates to the TTL values. To maintain consistency trainers follow the invariant discussed in \S\ref{sec:cache_prefetch_bp}.

%  \ie at the end of each batch the gradients of the dense neural networks are synchronized. 

%. The trainers perform prefetching by sending asynchronous requests to \embs.
% two other critical operations
%Cache maintenance . 

\noindent\textbf{\embs.} Embedding servers store all the embedding tables and act as a sharded parameter server, handling the prefetch and update requests from the trainers. 
% Our design is similar to prior DLRM designs~\cite{eisenman2022check} and thus 
We use the techniques presented in prior work~\cite{eisenman2022check} to checkpoint embedding servers periodically.
\subsection{Discussion}
% \vspace{-5pt}
\label{sec:discussion}
Next, we discuss some benefits and limitations of our design.

%\shivaram{Is a bit long}

\paragraph{Generalizing across skew patterns.}
Unlike prior work~\cite{adnan2021high}, \system's optimizations are resistant to embedding access skew changes (evaluated in \S\ref{sec:oracle_cacher_benchmark}). This is because \system does not just rely on caching of a fixed set of hot embeddings, it speeds up access to cold embeddings using pre-fetching. So if there exists datasets that do not display a high degree of skew, \system will still outperform prior work.
%but it should still be able to provide gains in throughput.
% just
% to provide speedups

\noindent\textbf{Applicability in online training.} \system's optimizations are applicable to offline setup, as it relies on the ability to look at future batches to build a cache. 
In case of online training examples arrive sporadically thus restricting lookahead. 

% oracular
%  which is guaranteed to provide a hit.

% \ziyi{can we suggest how our ideas could be (partially) applied to online training?}

\noindent\textbf{Scalability of \orc.} 
The overhead of \orc is extremely small even for extremely large batch sizes and lookahead values. In \S\ref{sec:oracle_cacher_benchmark} we find that \orc, even for large batch size of 131K, can dispatch 3.27 Million samples per second. Further, \orc only needs to be faster than the time taken by trainers for the forward and backward pass.
However, if required, \orc can be partitioned to increase scalability for datasets with a large number of embedding tables. To split the work done by \orc, we can partition the embedding tables such that each partition of the \orc can work on a different embedding table. For instance, if there are 1000 categorical features and we launch 10 \orc; for each example, each \orc generates caching decisions for their subset of 100 categorical features.
% \input{discussion}
% \vspace{-5pt}
\section{Implementation} 
% \vspace{-5pt}
 \system{} is implemented in around $5000$ lines of Python.  Async RPC's are used to communicate across different components. For synchronization of dense parameters and caches we use collective communication primitives present in NCCL \cite{nccl}. \system is completely integrated with PyTorch and  existing model training code can use it with 4 to 5 lines of changes. API details will be present in our open source version.

% As described previously, a
% . The embedding update request contains 

\noindent\textbf{Overlapping cache management with training.} 
We perform, all cache management operations in a separate thread thus not affecting the training process. Our caching data structure can operate completely lock free, because in our \orc's lookahead formulation, we guarantee that the training thread and cache maintenance thread will operate on completely separate indices of the cache. This ensures that cache management has minimal overhead on training.

\noindent\textbf{Automatically Calculating Lookahead.}
\label{sec:bpipe_config}
\system uses two configuration parameters: max cache size and lookahead value (\lkval). Providing the max cache size is mandatory, 
% If the user does not provide  
% \lkval can be automatically computed by \system. 
it can be determined by computing amount of free memory available after allocating space for the dense neural network parameters. \lkval can be automatically calculated if it is missing. To calculate \lkval, at startup \system keeps prefetching until it detects the cache is full. On detecting that the cache is full, \system{} selects the number of batches prefetched so far as the \lkval. Further, \system{} can also handle scenarios where the configuration variables are incompatible. Since \orc always has a consistent view of the cache, if it observes that the cache is going to be full it can reduce the \lkval. We perform a sensitivity analysis on \lkval in \S\ref{sec:oracle_cacher_benchmark}.

% \vspace{-5pt}
\section{Evaluation}
% \vspace{-5pt}
\begin{figure*}[t]
\centering
\begin{minipage}[t]{0.3\textwidth}
\includegraphics[width=0.9\textwidth]{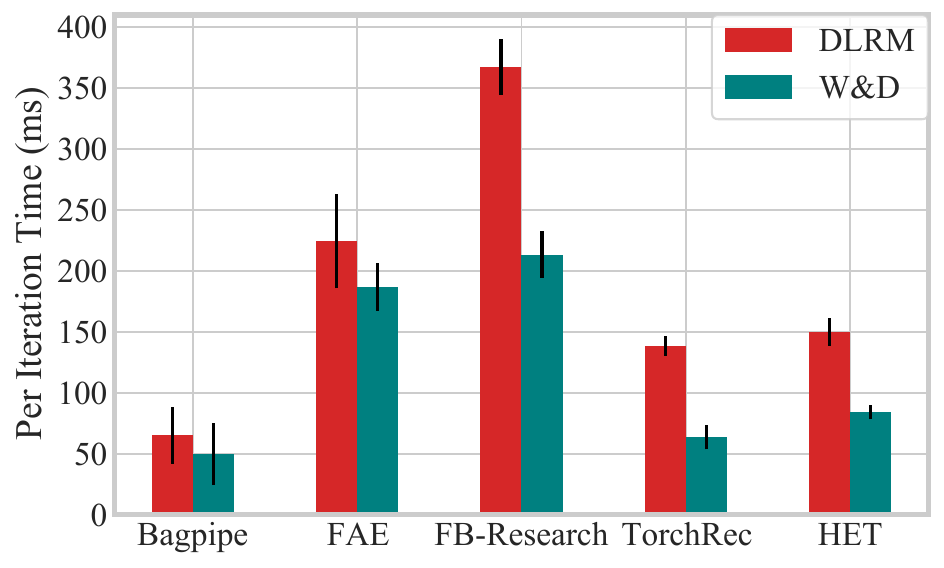}
\vspace{-0.2in}
\caption{\small{\textbf{Compare \system with Existing Systems}: We compare per iteration time of \system against existing \fae~\cite{adnan2021high}, FB-Research training system~\cite{dlrmopensource}, TorchRec~\cite{torchrec} and \het~\cite{miao2021het}. \system provides speedups betwen $1.2\times$ and $5.6\times$.}}
    \label{fig:compare_different_systems}
\end{minipage}\quad
\begin{minipage}[t]{0.3\textwidth}
\includegraphics[width=0.9\textwidth]{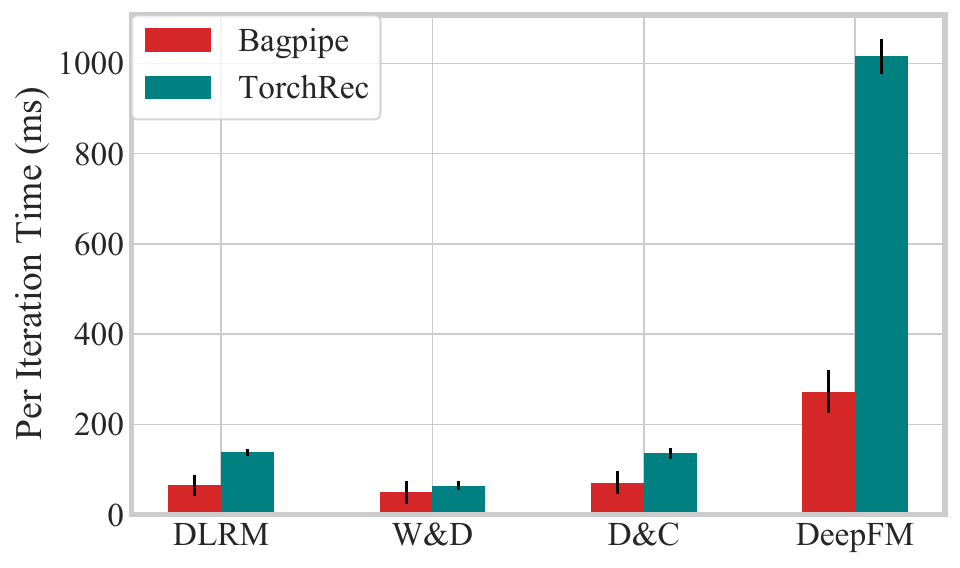}
 \vspace{-0.2in}
    \caption{\small{\textbf{Compare \system with different models}: We compare \system and \torchrec on four different models, DLRM~\cite{naumov2019deep}, W\&D~\cite{cheng2016wide}, D\&C~\cite{wang2017deep}, and DeepFM~\cite{guo2017deepfm}. We observe speedups between $1.2\times$ and $3.7\times$.}}
    \label{fig:compare_different_models}
\end{minipage}\quad
\begin{minipage}[t]{0.3\textwidth}
 \includegraphics[width=0.9\linewidth]{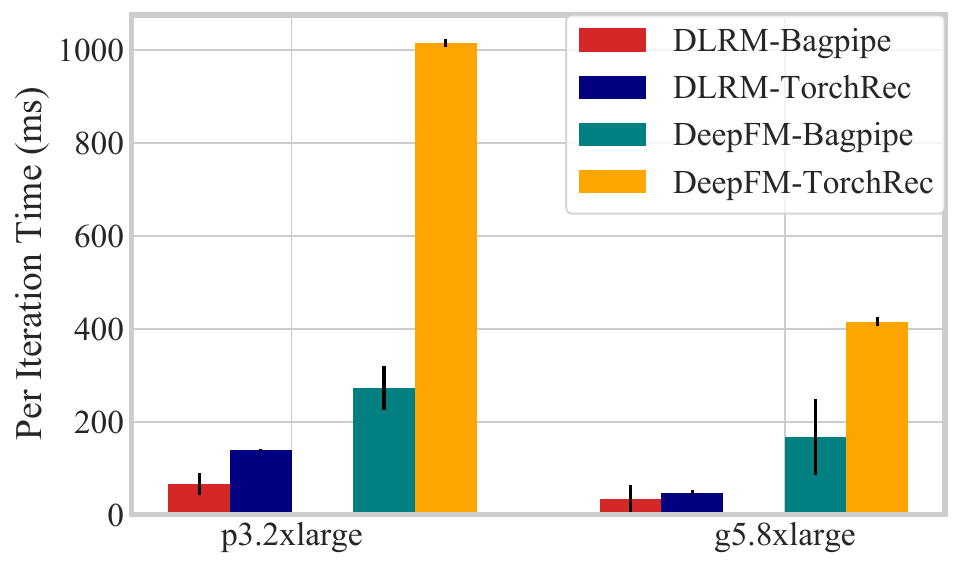}
    \vspace{-0.2in}
    \caption{\small{\textbf{Compare \system on different Hardware}: Speedup provided by \system over \torchrec on \emph{p3.2xlarge} decreases from $3.7\times$ to $2.5\times$ on \emph{g5.8xlarge} (high bandwidth) depicting that \torchrec is more constrained by bandwidth.}}
    \label{fig:compare_different_hardwares}\quad
\end{minipage}
\vspace{-17pt}
\end{figure*}

\begin{figure*}[t]
    \centering
    \begin{minipage}[t]{0.3\textwidth}
 \includegraphics[width=0.9\linewidth]{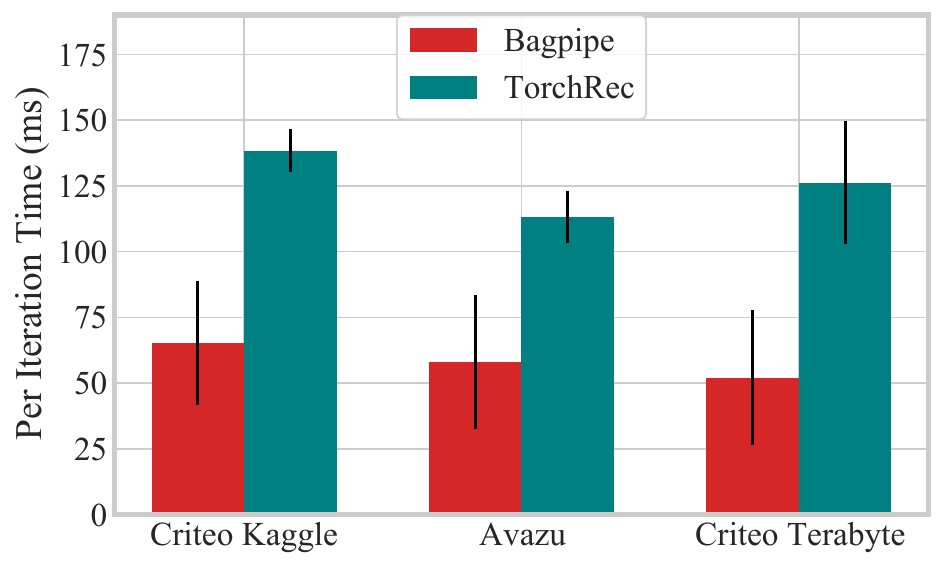}
    \vspace{-0.2in}
    \caption{\small{\textbf{Compare with Different Datasets}: \system consistently provides speedups between $1.9\times$ to $2.4\times$ across datasets. }}
    \label{fig:compare_different_datasets}
\end{minipage}\quad
    \begin{minipage}[t]{0.3\textwidth}
     \includegraphics[width=0.9\linewidth]{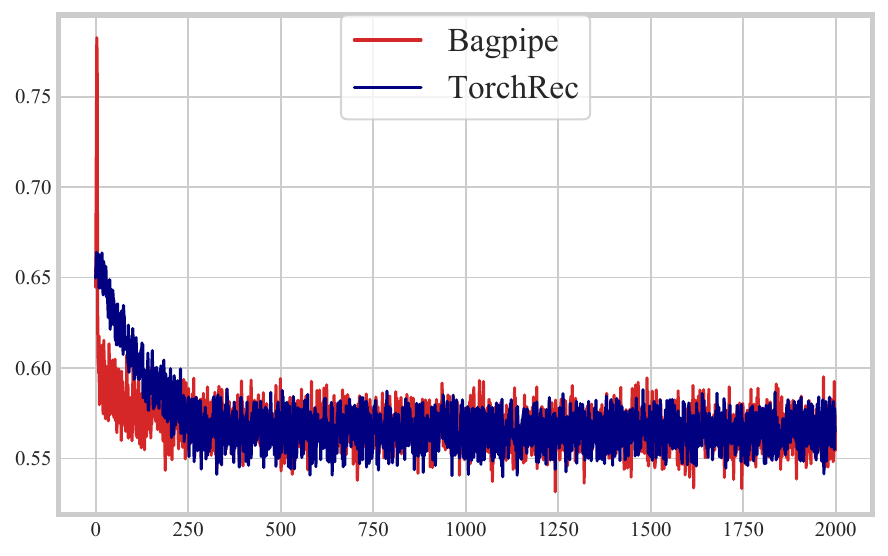}
    \vspace{-0.2in}
    \caption{\small{\textbf{Loss convergence for \system and \torchrec:} Convergence of \system{} and \torchrec is very similar, with slight differences  due to random initialization.}}
    \label{fig:loss_curve}
    \end{minipage}\quad
    \begin{minipage}[t]{0.3\textwidth}
    \includegraphics[width=0.9\linewidth]{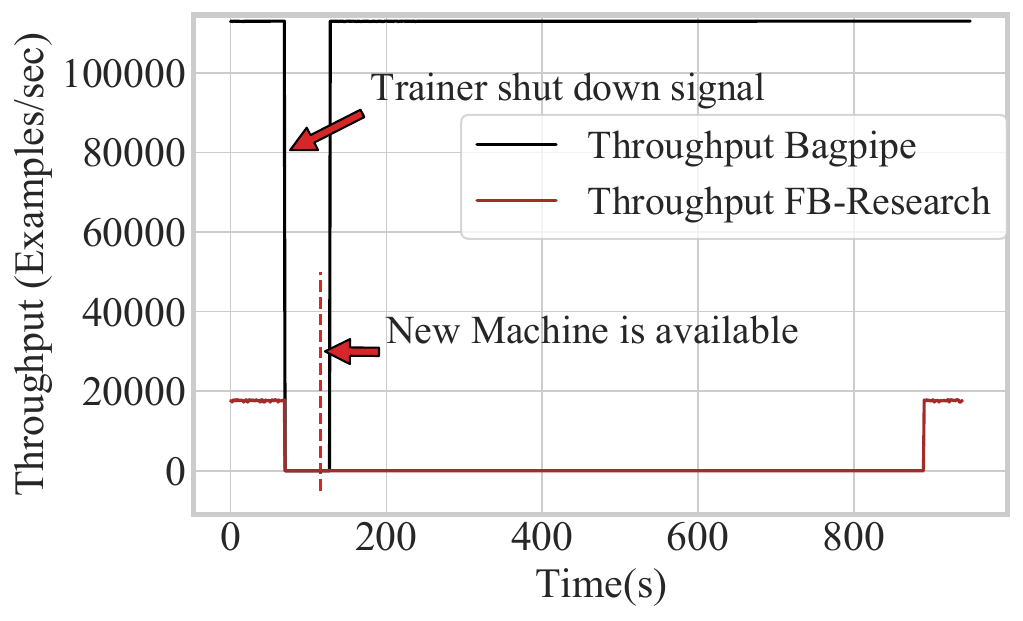}
    \vspace{-0.2in}
    \caption{\small{\textbf{Recovery from trainer failure}:} \system requires less than 60 seconds to recover from a trainer failure compared to 13 minutes for FB-Research System. 
    % This is because trainers in DLRM-Base hold a large amount of state leading to extremely slow recovery from checkpoints. 
    }
    \label{fig:checpoint}
    \end{minipage}\quad
    \vspace{-15pt}
\end{figure*}
% \vspace{-5pt}
\label{sec:eval}
We evaluate \system by 
measuring improvements in per iteration time against four baselines,  observing a speedup of $2.1\times$ to $5.6\times$
for the DLRM model. Further, we vary the recommendation model architecture and compare \system{} against the best-performing baseline with four different models and observe a speedup of up to $3.7\times$. We also analyze the performance of \system on different hardware and datasets and evaluate other aspects of \system like fault tolerance (\S\ref{sec:eval_ft}) and sensitivity to configuration parameters (\S\ref{sec:oracle_cacher_benchmark}).

\noindent\textbf{Baseline Systems.} To compare \system we use four open source baselines discussed in \S\ref{sec:existing_method}. We compare \system with FAE~\cite{adnan2021high}, FB-Research's training system~\cite{dlrmopensource}, TorchRec \cite{torchrec} and HET~\cite{miao2021het}. We discuss additional details of these systems when comparing them with \system in \S\ref{sec:existing_compare}.

\noindent\textbf{Models and Datasets.} We use four different recommendation models, Facebook's DLRM~\cite{naumov2019deep}, Google's Wide\&Deep~\cite{cheng2016wide}, Deep\&Cross Networks~\cite{wang2017deep}, and Huawei's DeepFM~\cite{guo2017deepfm}. Table~\ref{tab:models} describes the models used to evaluate \system. 
% and the number of dense parameters. 
The models   differ markedly in terms of the dense parameters, \eg the largest model has 33.8 Million parameters while the smallest one only has 136K parameters. 
For datasets, we use the Kaggle Criteo~\cite{kaggle}, Avazu~\cite{avazu} and Criteo Terabyte dataset~\cite{terabyte} (largest publicly available dataset). Table~\ref{tab:datasets} describes the embedding table size for each dataset.

\noindent\textbf{Cluster Setup.} We run all our experiments on Amazon Web Services(AWS). For trainers, we use \emph{p3.2xlarge} instances while \embs and \orc run on a \emph{c5.18xlarge} instance each.  Each \emph{p3.2xlarge} instance contains a Nvidia V100 GPU, 8 CPU cores and 64 GB of memory with inter-node bandwidth of up to 10 Gbps. Each \emph{c5.18xlarge} has 72 CPU cores and 144 GB of memory. For \system we launched dataloaders on the same \emph{c5.18xlarge} as \orc since the machine had ample compute. To study the performance of \system in a setting with different amounts of compute and bandwidth we also run some experiments on \emph{g5.8xlarge} where each machine has an Nvidia A10G GPU,32 CPU cores with inter-node bandwidth of 25 Gbps.

\begin{figure*}[t]
    \begin{center}
    \includegraphics[width=\textwidth]{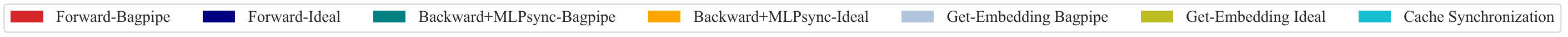}\\
    \vspace{-2pt}
    \begin{subfigure}[b]{0.24\textwidth}
        \includegraphics[width=\textwidth]{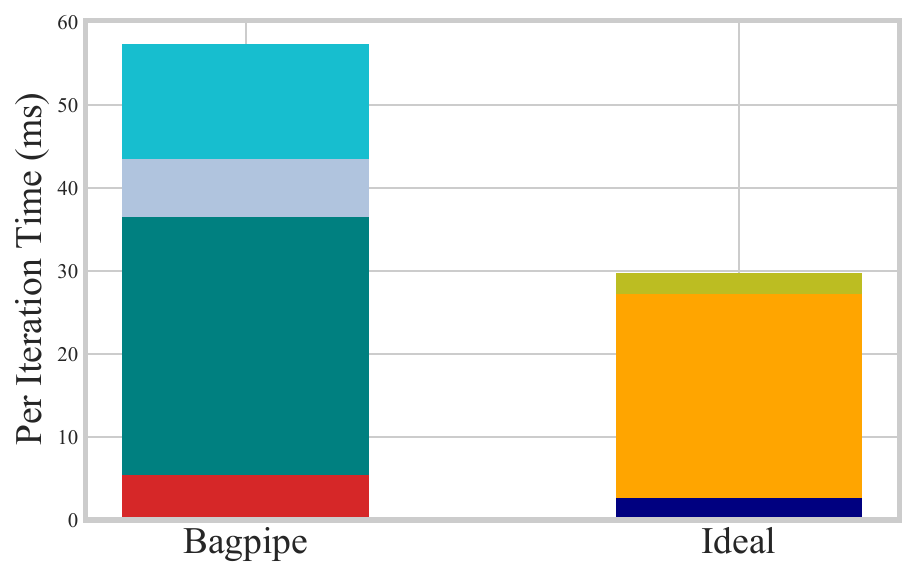}
        \vspace{-0.2in}
    \caption{DLRM: p3.2xlarge}
     \label{fig:p3_dlrm}
    \end{subfigure}
    \begin{subfigure}[b]{0.24\textwidth}
    \includegraphics[width=\textwidth]{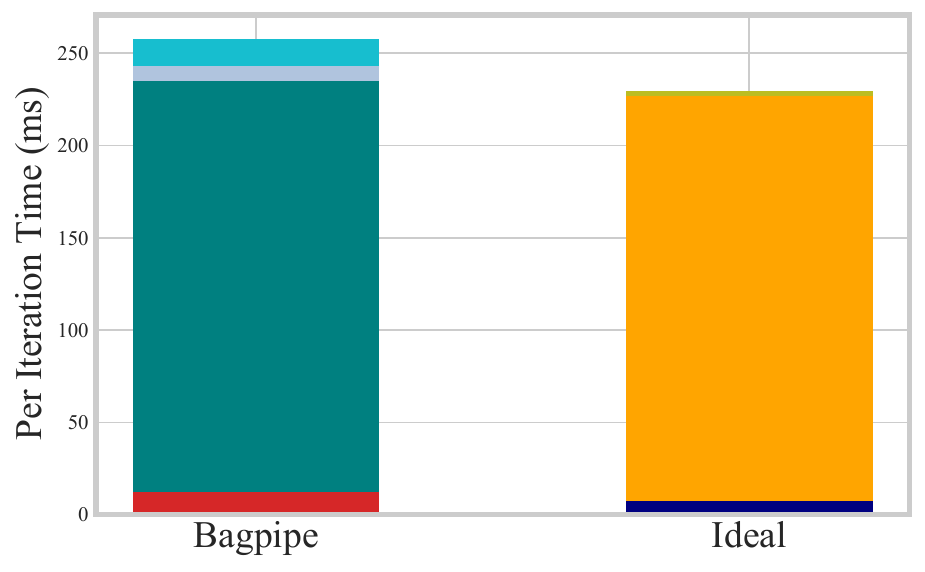}
    \vspace{-0.2in}
    \caption{DeepFM: p3.2xlarge}
    \label{fig:p3_deepfm}
    \end{subfigure}
    \begin{subfigure}[b]{0.24\textwidth}
    \includegraphics[width=\textwidth]{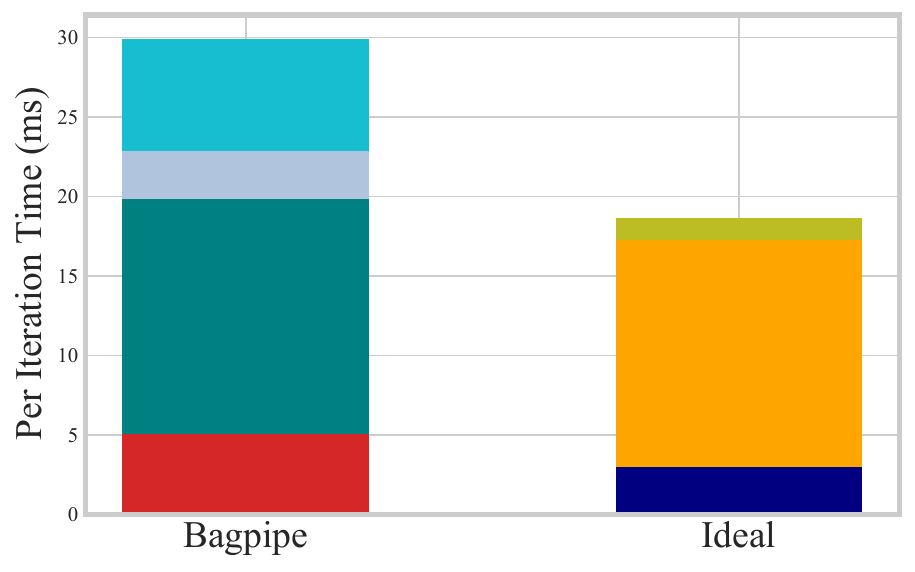}
    \vspace{-0.2in}
    \caption{DLRM: g5.8xlarge}
    \label{fig:g5_dlrm}
    \end{subfigure}
    \begin{subfigure}[b]{0.24\textwidth}
    \includegraphics[width=\textwidth]{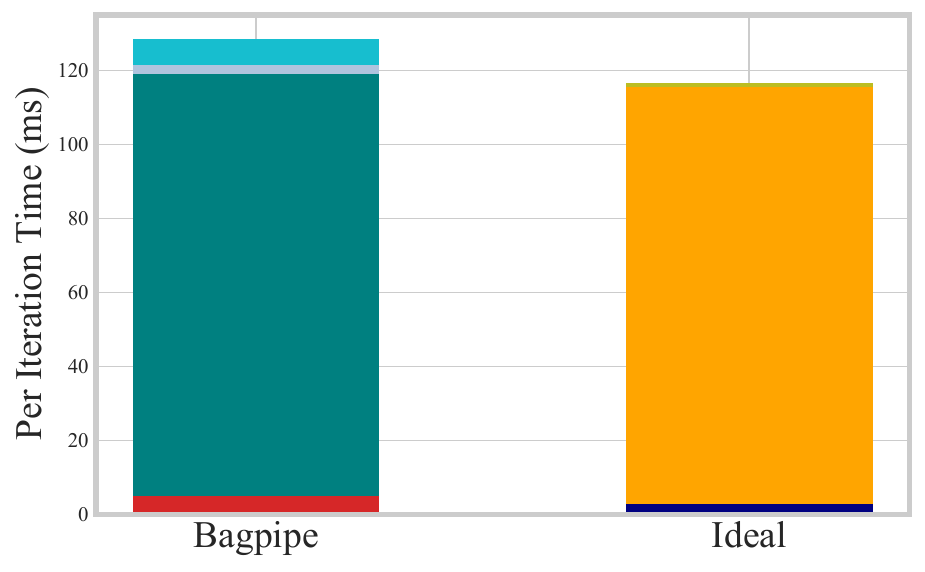}
    \vspace{-0.2in}
    \caption{DeepFM: g5.8xlarge}
    \label{fig:g5_deepfm}
    \end{subfigure}
    \end{center}
    \vspace{-0.2in}
    \caption{\small{\textbf{Comparing with Ideal:} Comparing \system with an \emph{ideal} system which has no overhead for embedding fetch, we observe that system comes within 10\% of time per iteration for large models where there is potential to overlap embedding accesses.}}
    \label{fig:comparing_with_ideal}
    \vspace{-15pt}
\end{figure*}

% \begin{figure}[t]
%     \centering
%         \includegraphics[width=0.4\textwidth]{images/scaling_comparisons_e2e.pdf}
%     \vspace{-0.1in}
%     \caption{\small{\textbf{Scalability of \system }: We perform both weak and strong scaling on \system. As we scale the benefits of \system get higher. For 8GPUs, \system provides a speedup of around $6\times$ over DLRM-Base while for 16 GPUs, it becomes around $7\times$. We use Criteo Kaggle dataset for these experiments.}}
%     \label{fig:scalability}
% \end{figure}

% We change the number of machines and batch size to evaluate scalability. 

\noindent\textbf{\system Configuration.} Unless otherwise stated, for all our experiments we set the cache size 
% such that we can 
% perform a lookahead of up to 200 batches. 
to enable lookahead of up to 200 batches.
We study the sensitivity of these parameters and their effect on throughput in \S\ref{sec:oracle_cacher_benchmark}. We run all our experiments for 2000 iterations, which roughly translates to 1 epoch of Criteo Kaggle Dataset with batch size of 16,384. 
% and memory

\noindent\textbf{Metrics.} For all our experiments we plot average per-iteration time with error bars representing standard deviation. This directly translates to the time taken to train a fixed number of epochs. As \system{} guarantees consistent access to embeddings, the accuracy after each iteration exactly matches other synchronous training baselines (validated in \S\ref{sec:existing_compare}).
%\begin{figure*}[t]   
%\end{figure*}
% \begin{figure}[t]
% \begin{minipage}[b]{0.32\textwidth}

% \end{figure}
% \end{minipage}\quad
%     \begin{minipage}[b]{0.66\textwidth}
%     \begin{subfigure}[b]{0.5\textwidth}
%     \includegraphics[width=0.8\textwidth]{images/diff_lookahead_cache.pdf}
%     \vspace{-5pt}
%     \caption{\small{\textbf{Cache Size}}}
%     \label{fig:effect_of_lookahead_cache_size}
%       \end{subfigure}
%         \begin{subfigure}[b]{0.5\textwidth}
%     \includegraphics[width=0.8\textwidth]{images/diff_lookahead_throughput.pdf}
%     \vspace{-5pt}
%     \caption{\small{\textbf{Throughput}}}
%     \label{fig:effect_of_lookahead_throughput}
%         \end{subfigure}
%     \vspace{-25pt}
%     \caption{\small{\textbf{Effect of increasing \lkval}: We study how changing \lkval effects Cache Size and Throughput. We observe that with increase in \lkval the cache size required also increases, since we need the cache space to store the extra embeddings from additional lookahead. However, the gain in throughput plateaus after \lkval of 100.}}
%     \label{fig:all_lkval_effect}
%     \end{minipage}
%     \vspace{-20pt}
% \vspace{-5pt}
\subsection{Comparing \system}
% \vspace{-5pt}
\begin{figure}[t]
    \begin{center}
    \begin{subfigure}[t]{0.23\textwidth}
       \includegraphics[width=1.0\linewidth]{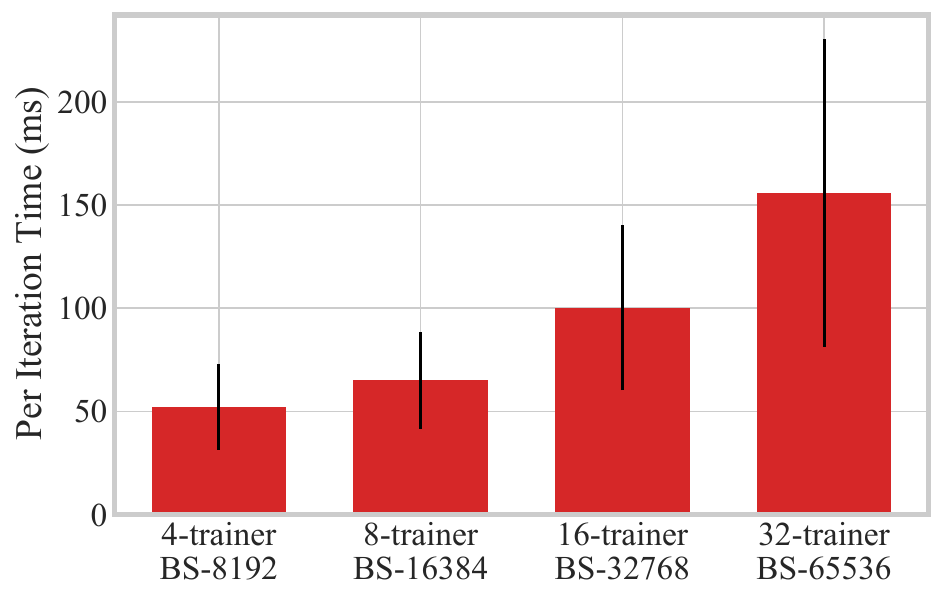}
    \vspace{-0.25in}
    \caption{\small{\textbf{Increasing Trainers} }}
    \label{fig:trainer_scalability}
    \end{subfigure}
    \begin{subfigure}[t]{0.23\textwidth}
    \includegraphics[width=1.0\linewidth]{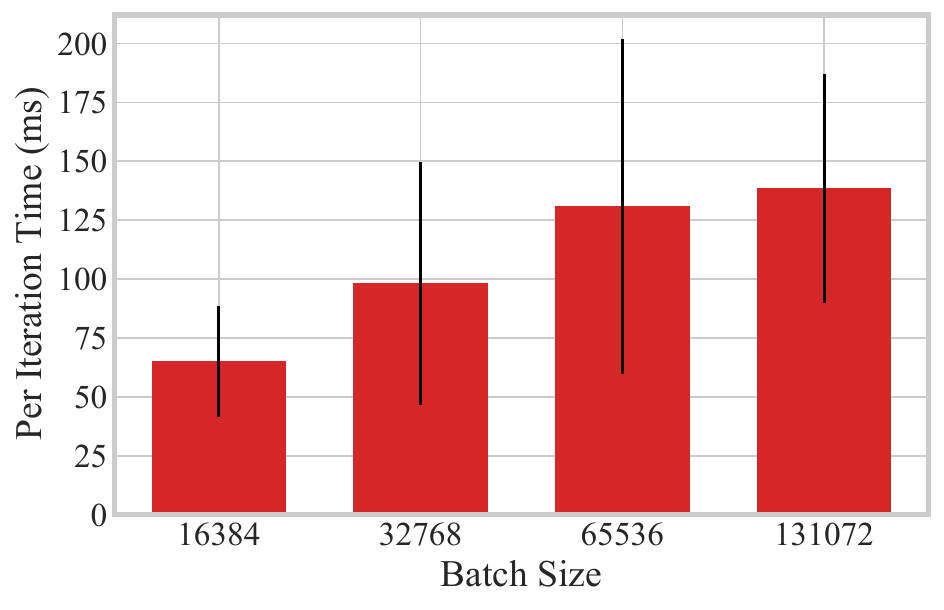}
    \vspace{-0.25in}
    \caption{\small{\textbf{Increasing Batch Size}}}
    \label{fig:scaling_batch_size}
    \end{subfigure}
    \end{center}
    \vspace{-0.2in}
    \caption{\small{\textbf{Scalability of \system:} \textbf{(left)} we increase the number of trainers such that batch size per machine is constant; \system provides sublinear scalability due to increasing communication bottlenecks. \textbf{(right)} Increasing batch size with 8 trainers results in better throughput as we are able to better overlap communication. }}
    \label{fig:scaling_comparison}
    \vspace{-10pt}
\end{figure}
\label{sec:existing_compare}
We first evaluate \system by comparing it against a number of existing systems and study how our benefits change as we vary the models, datasets, and hardware. 

\noindent\textbf{Comparing \system with existing systems.}
In Figure~\ref{fig:compare_different_systems}, we compare \system with four existing systems, \fae ~\cite{adnan2021high}, FB-Research training system~\cite{dlrmopensource}, \torchrec~\cite{torchrec} and \het~\cite{miao2021het}. We use Criteo Kaggle dataset with batch size 16,384 (a common batch size among MLPerf~\cite{mattson2020mlperf} entries) and two popular recommendation models DLRM~\cite{naumov2019deep} and W\&D~\cite{cheng2016wide}. 
% open source DLRM implementation by Facebook research~\cite{dlrmopensource} \shivaram{lets call this FB-Research instead of DLRM to avoid confusion with the model name?}, TorchRec~\cite{torchrec} and HET~\cite{miao2021het} \shivaram{Is it HET or HETU or Hetu? We use all three here..}.

\fae performs pre-processing on training data to classify embeddings as either hot or cold.
%It keeps hot embeddings in the GPU memory of the trainer. 
% FAE keeps the cold embeddings on the CPU while keeping the hot embeddings in the GPU memory. To avoid long tail of accesses FAE also partitions input data as either hot batches or cold batches, hot batches access embeddings only present in the GPU and cold batches may access embedding present on the CPU. To minimize synchronization between CPU and GPU and achieve maximum speedup, \fae prefers to train on all batches which only access hot embeddings first and then switch to batches with cold embeddings. However, this could lead to loss in statistical efficiency, to reduce this \fae uses additional logic to intertwine execution of cold and hot batches.
% \shivaram{this can go to S2}
%Since, the author provided \fae code is not compatible with a multi-node setup we implement FAE in our system while we use the author provided code to identify hot or cold embeddings. 
To evaluate the best case scenario for \fae, we do not account for the additional time \fae spends in partitioning batches and deciding the placement of embeddings. As shown in Figure~\ref{fig:compare_different_systems}, \system achieves $3.4\times$ speedups for the DLRM model and $3.7\times$ speedups for W\&D. 
As discussed in \S\ref{sec:existing_method}, during hot batch training \fae has similar cache synchronization overheads as \system, but when it switches to cold batches, it suffers additional embedding access overheads due to no prefetching.
% for that switch
% . \fae also incurs overheads when training with cold batches because unlike \system
%  FAE needs to switch between hot and cold batches to maintain accuracy and  this leads to significant slowdowns.
% \shivaram{was too long. tried to make it succinct.}

%The primary reason why \fae is slower than \system is that it spends a significant amount to synchronize embeddings when switching between hot and cold batches and during training of cold batches it provides no speedups over baseline \shivaram{there is no baseline. it suffers from increased synchronization times?}. Because of the additional overhead of up to 3 minutes when switching between hot and cold batches it ends being slower than \system.  
%To summarize,  

% Facebook research has open sourced a full system for synchronous training of deep learning based recommendation model.  the original system

Next, we compare \system with open source FB-Research training system~\cite{dlrmopensource}, built over PyTorch and Caffe-2, is designed for DLRM models~\cite{naumov2019deep} but can be easily modified to support other embedding-based deep recommendation models like W\&D~\cite{cheng2016wide}. \system provides $5.6\times$ and $4.2\times$ speedups over FB-Research training system. FB-Research system is slow due to spending almost 60\%  of the time on data loading, which has also been observed by prior works~\cite{zhao2021understanding, dataloaderissue} as well, leading to worse throughput compared to \system which offloads data-preprocessing to remote machines.
%to other machines. 

When compared against \torchrec, a recent open source system built over PyTorch~\cite{li2020pytorch} and FBGEMM~\cite{fbgemm} to facilitate training of recommendation of models, we observe that \system is around $2.1\times$ faster for DLRM models and around $1.3\times$ faster for W\&D models.
% \torchrec uses the highly optimized FBGEMM library for performing matrix operations. 
%Further, \torchrec tries to overlap independent operations to reduce observed time during training. \chengpo{These two 'further's look weird}
% During forward pass \torchrec overlaps embedding fetches with compute of the dense feature processing. Similarly during backward pass \torchrec overlaps backward compute with remote embedding writeback and synchronization of dense NN parameters. 
%Further, it also performs overlapping of training data movement to the GPU with model training. 
Unlike \system, \torchrec does not perform any caching or pre-fetching, and therefore fetches and writes back a large number of embeddings on the critical path. \system reduces and overlaps the amount of embedding-related communication on the critical path. 
% on the critical path
% \shivaram{Add a sentence here of what do they not do which we do?}

%  and heavily optimized

To compare with \het~\cite{miao2021het}, a system that performs bounded asynchronous training as described in \S\ref{sec:existing_method},
%In this regime, the model can be trained with stale embeddings within a certain bound. 
%To reduce embedding fetch overhead, \het caches most frequently used embeddings in a local cache present on the workers. 
% however, unlike \system the caches are not synchronized and stale embeddings can be cached as long as they are within the configured asynchrony bound. 
%If they are beyond the asynchrony bound, \het synchronizes these embeddings with the embedding server using them. \chengpo{Use parameter server for HET consistently?} 
we use the author-provided code implemented in C++ with Python bindings to evaluate \het and set the asynchrony bound to 100, as suggested by the authors for maximum speedup. 
We find that \system is around $2.3\times$ faster than \het for DLRM and $1.6\times$ faster for W\&D. We observe that, despite performing asynchronous training, \het needs to fetch embeddings that are not available in the local cache from the parameter server on the critical path. With increase in batch size the number of cache misses increases as well, due to the long tail of accesses (discussed in \S\ref{sec:access_patterns}). 
We also verify that our performance closely matches with those reported in the paper~\cite{miao2021het}.
%  which instead of performing synchronous training, uses a
% \het is optimized to overlap operations like cache evictions, embedding fetches with the the compute. 

% Further, we use 8 p3.2xlarge machines for trainings. HET~\cite{miao2021het} and \system require additional parameters servers, for those we use c5.8xlarge instances. For HET we used the asynchrony bound of 100 which authors have shown provides maximum speedup without sacrificing accuracy. For HET, our obtained numbers also closely match with those reported in the original paper, hence verifying that we are indeed using the correct configuration.
% As shown in Figure~\ref{fig:compare_different_systems} we observe that \system outperforms every system on both DLRM and W\&D models. We observe for DLRM model \system provides a speedup of $3.4\times$ over FAE, $5.6\times$ over FB-Research, $2.1\times$ over TorchRec and $2.2\times$ over HET.  
% %Restart writing here

As the speedup of \system varies across models, we perform a detailed investigation to understand this. 
We observe \torchrec to be the best performing baseline as it efficiently overlaps different parts of the training pipeline and make better use of network bandwidth using the \texttt{all2all} primitive for embedding fetches.
 %, unlike \het which depends on a parameter server.
 %We also observe that the speedup provided by \system can depend significantly on the structure of the model, \eg, for DLRM, \system provides a speedup of over $2.1\times$ over \torchrec but for W\&D we observe only $1.3\times$ speedup. 
 % Since \torchrec is the best performing baseline, 
 Thus, for the next set of experiments we compare \system with \torchrec.

% FAE on the other suffers due to cache synchronization required in the \shivaram{Incomplete sentence. Didn't introduce FAE before.}

% In summary, we observe that \system outperforms existing synchronous as well as asynchronous training by a wide margin without sacrificing the guarantees of synchronous training. 

%\shivaram{It would be good to show a breakdown of at least one comparison. Say TorchRec to show where the wins are coming from. }

%\shivaram{Also would be good to say why wins are more in DLRM? If you want to save that for next section then maybe we dont need W\&D here?}

% Additionally, to delineate the benefits due to implementation and understand the gains provided by by  caching and pre-fetching we run \system without using caching and prefetching as well.
% For these experiments we use Criteo Kaggle Dataset and 8 Trainer machines.
% We also fix the batch size as 16,384.

% \todo{Add a figure here comparing these systems. Let's do per iteration times here, and split it in two heads communication and computation, the figure will be a bar graph with Systems on X-axis, do like DLRM Base, TorchRec, HET, Bagpipe-No cache No prefetch, Bagpipe}

\noindent\textbf{Comparing \system on other models.}
%Previously we compared \system on DLRM and W\&D models. Next, we study how do speedups provided by \system change with changes in model architecture. For these comparisons we used \torchrec since we observe that it is the best performing baseline.
% \shivaram{We already have a bridge in end of prev para.}
In addition to W\&D and DLRM, we also train the Deep\&Cross Network (D\&C)~\cite{wang2017deep} and DeepFM models~\cite{guo2017deepfm} with \system and \torchrec. D\&C models contain an additional Cross Network component, which performs explicit feature crossing of sparse features to learn predictive features of bounded degree without manual feature engineering. DeepFM introduces a factorization module that learns up to 2-order feature interactions between sparse and dense features. 
% These models increase the number of dense parameters compared to W\&D (136K vs 22.5 Million DeepFM) thus requiring more bandwidth during synchronization.
% We would like to note that all these models only change the structure and number of parameters in the dense portion of the recommendation models and do not effect the embedding access pattern or the size of embedding tables. 
Details of these models are available in Table~\ref{tab:models}.
In Figure~\ref{fig:compare_different_models} we observe that performance gains provided by \system over \torchrec depends on the size and computation requirements of the dense portion of recommendation models, \eg for W\&D which only has around 131,000 dense parameters we observe that \system provides only a $1.2\times$ speedup, while for DeepFM which has 33.8 million parameters \system provides a speedup of over $3.7\times$. 
We believe that this is due to the pipelining mechanism present in \torchrec where the authors overlap the embedding write-back with the synchronization of the dense model. % to hide latency of embedding reads and writes. 
%The one crucial overlap mechanism which \torchrec has,
As the model size increases, the bandwidth requirement for synchronization also increases and the synchronization of dense model and embedding write-backs ends up competing for the same set of network resources. Meanwhile, \system significantly reduces the amount of embedding synchronization due to caching. Further, delayed  synchronization allows \system overlap forward pass of the next iteration. 
Thus, our analysis indicates that \torchrec, unlike \system, is heavily bottlenecked by the network bandwidth available. To verify this, we next run experiments on a different hardware. 

% Thus for a large network it is not possible to overlap embedding write back with model synchronization.

% In Figure~\ref{fig:compare_different_models} we observe that \system can improve per iteration times by upto $3.7\times$ for a parameter heavy model like DeepFM. The speedup is minimum for the small W\&D model because TorchRec is able to pipeline operations better as the synchronization time for the dense model is smaller. In general, we observe that TorchRec is highly susceptible to model size, as when the bandwidth requirement for the model increases and the ability to pipeline operations like embedding fetches significantly reduces. Unlike, TorchRec, \system reduces the amount of embedding fetches and is also able to overlap all the embedding fetches. 

% \shivaram{maybe this needs to be explained in S2 as to what Torchrec pipelines, but it is not very clear  by just reading this. }
\noindent\textbf{Comparing \system on different hardware.}
To understand the performance of \system on different hardware setups, especially in terms of network bandwidth, we evaluate \system on \emph{g5.8xlarge} (A10G GPU and 25 Gbps bandwidth) instances. In terms of compute, A10G  performs similar to V100, but the bandwidth on \emph{g5.8xlarge} is 25 Gbps compared to 10 Gbps on \emph{p3.2xlarge}.
% is significantly higher than \emph{p3.2xlarge} (25 Gbps vs 10 Gbps).
We use the same hyper-parameters and model configurations as in previous sections. Figure~\ref{fig:compare_different_hardwares} shows a comparison of per-iteration time between DLRM and DeepFM, for both \system and \torchrec. We observe that \system with DLRM model on \emph{g5.8xlarge} trainers is around $1.9\times$ faster than $p3.2xlarge$ trainers. 
% it takes around 34 ms 
% (around $1.9\times$ reduction). 
On other hand, the DLRM model with \torchrec on \emph{g5.8xlarge} trainers is around $2.4\times$ faster than \emph{p3.2xlarge} trainers.
% takes around 138ms per iteration while on \emph{g5.8xlarge} instances the per-iteration time is around 57ms ($2.4\times$ reduction). 
Similarly for DeepFM, time for \torchrec reduces by $2.4\times$ (1015ms to 414ms) when we switch from \emph{p3.2xlarge} instances to \emph{g5.8xlarge}. This confirms our hypothesis that for larger models \torchrec is bounded by bandwidth, while \system, because of caching and efficient pipelining of communication, makes better use of network resources. However, it is unclear yet, what fraction of the iteration time in \system is spent on network-bound embedding access and to understand this, we next compare \system{} to an \emph{ideal} system which has no overhead of embedding accesses.

%  compute bound, 
%\shivaram{need a slightly better takeaway and bridge.}
%This shows that as number of parameters increase, \system will still not be completely bottle-necked by network bandwidth unlike systems like \torchrec. 

% These results prompt us to study how much overhead does \system have due handling embeddings, \ie How will  \system compare to an \emph{ideal} system which has no overhead of embedding accesses ?

% \saurabh{rework this a bit, we want to convey that sys}
% that at higher bandwidths the speedup provided by \system reduces compared to TorchRec, from $2.1\times$ to $1.4\times$ for DLRM  and $3.7\times$ to $2.4\times$. \todo{Waiting for ideal}

\noindent\textbf{Comparing \system with an \emph{ideal} system.}
% Next we study how \system compares with an \emph{ideal} system which has no overhead to access remote embeddings.
%Next, we compare \system to an \emph{ideal} system. This 
Comparing \system{} to an \emph{ideal} system will show how far \system is from completely alleviating embedding access overheads. %\saurabh{Shivaram check}
% and how far \system's design has come in alleviating the embedding  \shivaram{add one sentence why this is interesting. Something like this shows us how effective our caching design is at minimizing overheads and how much room there is for further improvements.}
To create such an ideal system, we prefetch all necessary embeddings to the GPU memory before starting training and switch off prefetch, cache sync, and cache eviction modules of \system. In Figure~\ref{fig:comparing_with_ideal}, we perform this comparison for DLRM and DeepFM models on both \emph{p3.2xlarge} and \emph{g5.8xlarge}. DLRM model on \emph{p3.2xlarge} instance on \emph{ideal} system takes around 30ms while, \system takes around 56ms. DLRM model on \emph{g5.8xlarge} on the \emph{ideal} system takes around 19ms while \system takes around 30ms. This shows that at lower bandwidths for DLRM, there are periods in the pipeline when model training is blocked on embedding operations. We also study the same effect with DeepFM, a larger model that provides more opportunities for \system to overlap embedding-related operations. For DeepFM we observe that on \emph{p3.2xlarge} instances {ideal} takes around  236ms while \system takes around 253ms (overhead 17ms). On the high bandwidth \emph{g5.8xlarge} instance \emph{ideal} system takes around 116ms while \system takes around 128ms (overhead of 12ms). These results indicate that \system gets within 10\% of an \emph{ideal} system with deeper models and has almost constant overhead for providing embeddings (around 12 to 20 ms) even at lower bandwidths.

% which does not have any overhead for embedding related operations

\noindent\textbf{Comparison on Different Datasets.}
We also analyse performance of \system on % To analyze the performance of \system on da
Avazu~\cite{avazu} and Criteo Terabyte \cite{terabyte} (largest publicly available dataset). Using eight \emph{p3.2xlarge} instances as trainers, in Figure~\ref{fig:compare_different_datasets} we see that compared to \torchrec on DLRM model, \system is $1.9\times$ to $2.4\times$ faster. This shows that irrespective of the dataset \system provides a significant speedup over the best-performing baseline. 

\noindent\textbf{Convergence Comparison.}
%\shivaram{Lets move this up to the systems comparison section 5.1}
Since \system ensures that embedding reads are not stale, it should have the same convergence properties as synchronous training using \torchrec. We verify this in Figure~\ref{fig:loss_curve} where we see that \system's convergence is very close to \torchrec with minor differences arising from random initialization. For \het we observed that the convergence depends on the model complexity; for DLRM, \het's open source code~\cite{hetopensource} did not converge to the same loss as \torchrec, while we observed similar convergence as \torchrec for W\&D, a smaller model (We have reported this issue to the \het authors). Overall, we see that \system retains the convergence of synchronous training while providing per-iteration speedups.
\subsection{Scalability and Fault Tolerance}
% \cm{In following para the scalability of and fault tolerance of \system}

\label{sec:eval_ft}
% \vspace{-5pt}
\noindent\textbf{Scalability.} 
In Figure~\ref{fig:trainer_scalability}, we scale batch size and number of machines (up to 32 GPUs and batch size of 65,536). 
With increase in batch size the number of embeddings to fetch and synchronize increases. Despite increase in communication, \system scales around $1.4\times$ for $2\times$ increase in resources and work (320K samples/sec for 16 trainers vs 446K samples/sec for 32 trainers).
% We study the scalability of \system in two scenarios: increasing the batch size with corresponding increase in  number of machines and increasing batch size with a fixed number of machines.
% In Figure~\ref{fig:trainer_scalability}, we increase both the batch size and the number of machines (up to 32 GPUs and batch size of 65,536). As batch size increases, 
% the number of embeddings fetched and the number of cache elements synchronized increases. Despite the increase in amount of communication, \system scales around $1.4\times$ for $2\times$ increase in resources and work (320K samples/sec for 16 trainers vs 446K samples/sec for 32 trainers).
In Figure~\ref{fig:scaling_batch_size} we scale just batch size, we observe that batch size 65,536 takes a very similar time as batch size 131,072. Because with a higher batch size \system is able to overlap a bigger proportion of cache synchronization with the longer forward pass.
\begin{figure}[t]
    \begin{center}
    % \begin{subfigure}[b]{0.3\textwidth}
    % \includegraphics[width=0.8\linewidth]{images/oracle_cacher_lookahead.pdf}
    % \vspace{-0.1in}
    % \caption{\small{\textbf{\lkval}}}
    % \label{fig:effect_of_lookahead_throughput_oracle_cacher}
    %   \end{subfigure}
    \begin{subfigure}[t]{0.43\linewidth}
   \includegraphics[width=1.0\linewidth]{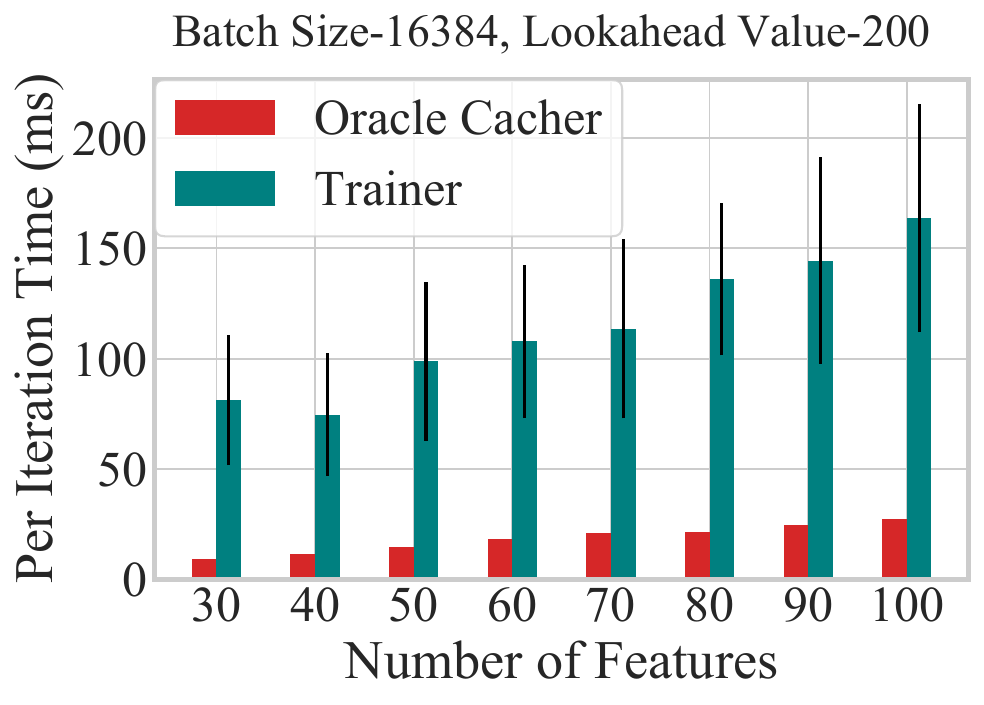}
    \vspace{-20pt}
    \caption{\small{\textbf{Categorical Features}}}
    \label{fig:effect_of_feat_throughput_oracle_cacher}
          \end{subfigure}
        \begin{subfigure}[t]{0.43\linewidth}
    \includegraphics[width=1.0\linewidth]{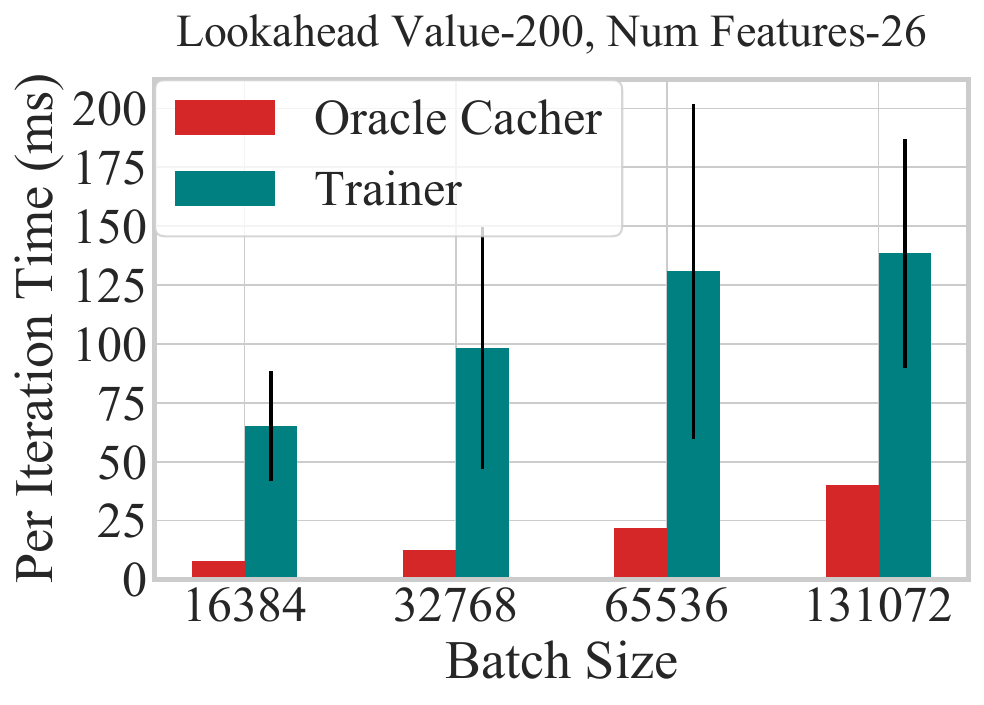}
    \vspace{-20pt}
    \caption{\small{\textbf{Batch Size}}}
    \label{fig:effect_of_bs_throughput_oracle_cacher}
        \end{subfigure}
    \end{center}
    \vspace{-10pt}
    \caption{\small{\textbf{Latency of \orc}: We observe that overall \orc scales very well, it increases sub-linearly with the increase in the number of features and batch size. However, training time will always hide the latency of \orc.}}
    % \vspace{-20pt}

\vspace{-10pt}
\end{figure}

\noindent\textbf{Fault Tolerance.}
In Figure~\ref{fig:checpoint} we observe that trainer in \system recover in less than a minute, compared to FB-research system which is close to 13 minutes. 
For the FB-Research  system we make a best-case assumption that the framework can checkpoint the iteration just before failure, to avoid checkpointing at every iteration. 
Even in this case, FB-Research  system takes around 13 minutes to recover since the amount of state on each trainer includes a large shard of the embeddings. Meanwhile, trainers in \system{} are able to recover in less than a minute and do not require checkpointing at every iteration (\S\ref{sec:ft}). Other systems do not discuss fault tolerance. 

\begin{table}[t]
\begin{center}
 \caption{\small{\textbf{Effect of increasing \lkval}: With increase in \lkval the cache size required increases but improves throughput till \lkval of 100.}}
 \vspace{-10pt}
 \label{tab:lkval_change}

\resizebox{0.6\linewidth}{!}{
\begin{tabular}{@{}|ccc|@{}}
\toprule
Lookahead                 & Cache Size (MB)                 & Avg Time per Iteration (ms) \\ \midrule
\multicolumn{1}{|c|}{5}   & \multicolumn{1}{c|}{39.6}   & 535.6                       \\
\multicolumn{1}{|c|}{10}  & \multicolumn{1}{c|}{66.7}   & 289.3                       \\
\multicolumn{1}{|c|}{50}  & \multicolumn{1}{c|}{235.2}  & 85.7                        \\
\multicolumn{1}{|c|}{100} & \multicolumn{1}{c|}{410.5}  & 67.3                        \\
\multicolumn{1}{|c|}{200} & \multicolumn{1}{c|}{720.6}  & 65.1                        \\
\multicolumn{1}{|c|}{300} & \multicolumn{1}{c|}{1003.3} & 65.6                      \\ \bottomrule
\end{tabular}}
\end{center}
\vspace{-15pt}
\end{table}
% \vspace{-5pt}
\subsection{Sensitivity Analysis of \system}
% \vspace{-5pt}
Next, we study the performance of \system with different configurations and also micro-benchmark components of \system. Unless stated otherwise, we use the same setup described \S\ref{sec:eval}, with a batch size of 16,384 on Criteo Kaggle dataset.

\noindent\textbf{Overhead of \orc.}
\label{sec:oracle_cacher_benchmark}
In Figure~\ref{fig:effect_of_feat_throughput_oracle_cacher},~\ref{fig:effect_of_bs_throughput_oracle_cacher} we observe  \orc's overhead
increases sub-linearly with increase in categorical features and batch size. However, the time per iteration is still significantly higher than the time taken to perform lookahead by \orc. Since \orc is overlaped with training, it only becomes a bottleneck if it's time exceeds that of trainer.
Overall, we find that \orc can almost dispatch \emph{3.27 Million} examples per second. We find that this is sufficient to power the most optimized systems reported in prior work (e.g., 8 ZionEx nodes (128 A100 GPUs) processing up to 1.6 Million samples per second~\cite{mudigere2021software}). We benchmarked \orc for other parameters like different \lkval and observe constant throughput, \ie complexity of \orc does not depend on \lkval.

\noindent\textbf{Effect of \lkval.} In Table~\ref{tab:lkval_change} we study how cache size required and throughout changes for different \lkval. As \lkval increases, cache size required increases sub-linearly. This sub-linear behavior due to reuse of embeddings found in the previous batches during lookahead process.
We also observe that throughput benefits from increasing \lkval start plateauing beyond 100. This is because as the lookahead value goes over 200, we are keeping all the popular elements in the cache, and increasing \lkval at this point does not affect communication much. 
% In Table~\ref{tab:lkval_change} we study the how cache size required 

\begin{figure}[t]
    \begin{center}
    \includegraphics[width=0.8\linewidth]{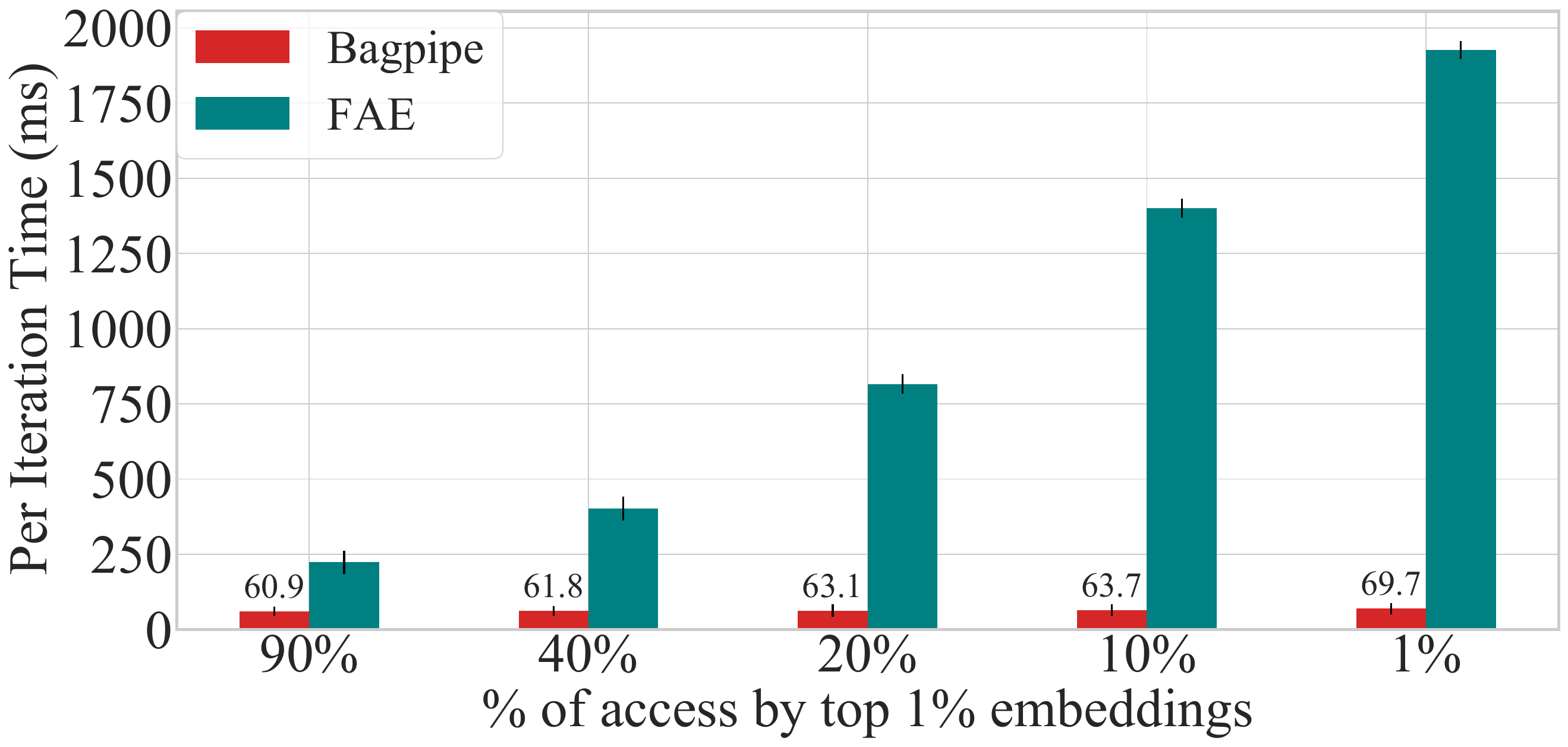}
    \vspace{-10pt}
    \caption{\small{\textbf{Effect of change in skew:} Comparing  when 1\% of embeddings perform 90\% of embedding accesses to  just 1\% of embedding access (no skew). Unlike \fae, \system's time only increases from 60.9ms to 69.7ms showing resistance to change in skew.}}
    \label{fig:different_hotness}
    \vspace{-15pt}
    \end{center}
    \end{figure}

\begin{figure}[t]
    % \begin{subfigure}[t]{\linewidth}
    \begin{center}
    \includegraphics[width=0.8\linewidth]{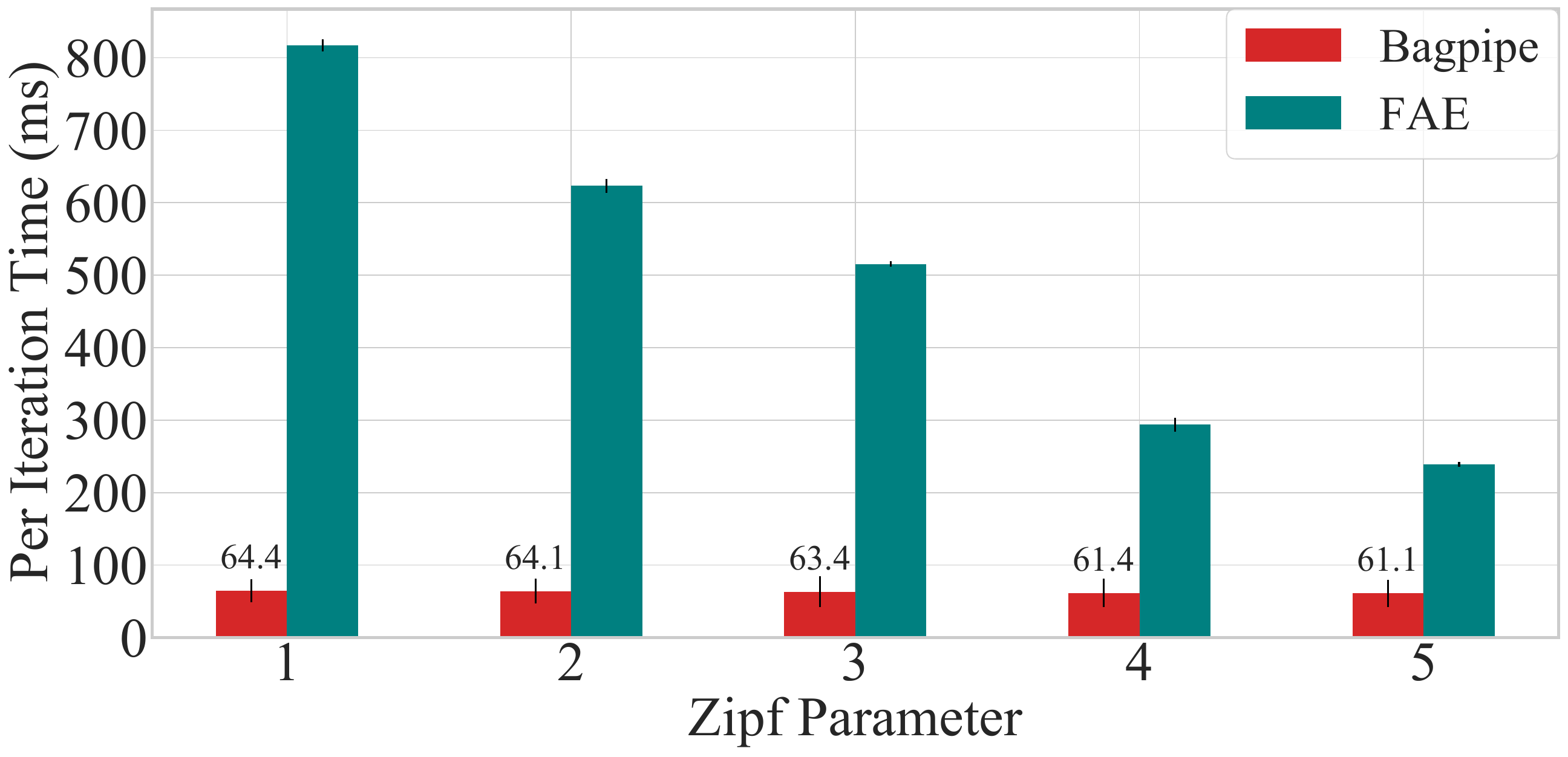}
    \vspace{-10pt}
    \caption{\small{ \textbf{Effect of change in skew using Zipf Distribution:} Varying the $\alpha$ parameter in Zipf distribution; a higher $\alpha$ indicates higher skew. Even with drastic increase in the skew, the time taken by \system remains almost constant.}}
    \label{fig:different_hotness_zipf}
    \vspace{-15pt}

    \end{center}
    \end{figure}

\noindent\textbf{Effect of Access Pattern Skew.} Unlike some prior systems~\cite{adnan2021high}, \system is designed to handle skew pattern changes. To study performance of \system when the skew pattern changes, we create an artificial dataset similar to Criteo Kaggle dataset with the same number of features and samples but with different skew patterns. 
We choose top 1\% of embeddings and then create an exponential function such that cumulative probability of sampling from top 1\% embeddings is equivalent to the chosen skew, \eg top 1\% of embeddings are responsible for 40\% accesses. The remaining embeddings are sampled uniformly such that they lead to the remaining 60\% of accesses. 
In Figure~\ref{fig:different_hotness} we study how the iteration time changes as the embedding reuse of top 1\% embeddings changes between 90\% and 1\%; \eg the 40\% bar reflects the runtime when 1\% of embeddings are reused 40\% of the time. 
We observe that due to optimizations present in \system like pre-fetching, LRPP and delayed synchronization even when the degree of skew changes from 90\% skew to no skew, \system's per iteration time changes at most by 13\%. On the other hand, FAE~\cite{adnan2021high}, which relies only on caching degrades by $7.2\times$. Next we vary the skew of embedding accesses using the popular Zipf~\cite{kingsley1932selected} distribution. The $\alpha$ parameter in Zipf distribution determines the skew, with a higher $\alpha$ denoting higher skew.  In Figure~\ref{fig:different_hotness_zipf} we observe that even with a large change in skew (varying Zipf's parameter between 1 and 5), \system's throughput does not vary significantly. This shows \system is resistant to changes in skew.

% \vspace{-5pt}
\section{Related Work}
 % \vspace{-5pt}
Multiple systems have been developed for training large DNN's. Systems like PipeDream~\cite{narayanan2019pipedream} and Gpipe~\cite{huang2019gpipe} introduce pipeline parallelism to enable efficient model parallel training. These systems are designed for dense training, while recommendation models have sparse access patterns, also requiring a hybrid data and model-parallel setup which is not supported in these systems.

Prior domain specific systems like Marius~\cite{mohoney2021marius} and P3~\cite{gandhi2021p3} for Graph Neural Network training, Oort~\cite{lai2021oort} and FedML~\cite{chaoyanghe2020fedml} for federated learning do not support offline recommendation model training. 

Pipeswitch~\cite{bai2020pipeswitch} and SwitchFlow~\cite{wu2021switchflow}  build mechanisms for fast preemption to schedule training and inference jobs to improve resource utilization. \cite{bai2020pipeswitch} relies on the idea that model weights have a deterministic access pattern (a simple example for PipeSwitch is that Layer 1 will be accessed before Layer 2), and uses this to start computation before moving the full model to the GPU. Unlike \system, Pipeswitch does not handle cases where the model weights (embeddings) are accessed dynamically depending on the training data.  

For offline training of recommendation models several prior systems like \fae~\cite{adnan2021high}, \torchrec~\cite{torchrec}, \het~\cite{miao2021het} have been designed. We have compared to these systems in the \S\ref{sec:eval} and discussed how \system is different. Prior work~\cite{acun2021understanding, liu2022monolith} has also proposed system designs to enable disaggregated training of recommendation models. Unlike \system, prior systems (e.g., Monolith~\cite{liu2022monolith}) suffer from embedding access overheads. There have been other systems like cDLRM~\cite{balasubramanian2021cdlrm} and ScratchPipe~\cite{kwon2022training} which have used the idea of lookahead to provide access to embeddings. However, these systems are primarily designed for single node training. Scaling these systems to multi-node setting  is non-trivial ~\cite{kwon2022training} or will impose a restriction on users to use specific optimizers like~\cite{balasubramanian2021cdlrm}. Systems like cDLRM~\cite{balasubramanian2021cdlrm} only allow embedding averaging rather than gradient averaging, which restricts users from using optimizers like Adam~\cite{kingma2014adam} and SGD with Momentum~\cite{nesterov2018lectures} thereby potentially harming convergence and changing the underlying training algorithm.  Other approaches like TTRec~\cite{yin2021ttrec} and~\cite{gupta2021training} improve performance using approximations like tensor compression and gradient compression, which unlike \system, change the training algorithm and can lead to accuracy loss.
 %\vspace{-5pt}

\section{Conclusion}
%\vspace{-5pt}
We presented \system{}, a new system that can accelerate the training of deep learning based recommendation models. Our gains are derived from better resource utilization and by overlapping computation with data movement. Our disaggregated architecture also allows independent scaling of resources and better fault tolerance, while retaining synchronous training semantics. Our experiments show that \system provides an end-to-end speedup of up to $5.6\times$ over state-of-the-art baselines.

\paragraph{Acknowledgements}
We would like to thank Carole-Jean Wu for introducing us to deep learning recommendation models and the challenges in scaling their training. 
We would also like to thank Yibo Zhu for early discussion about our work. 
Finally, we would like to thank our shepherd, Junfeng Yang, the anonymous SOSP reviewers, Bilge Acun-Uyan and Muhammad Adnan for their invaluable feedback that helped in making this work better. 
This research was supported in part by NSF-CAREER grant CNS-2237306.
% \section{TODO:}
% for saurabh to keep track
% \clearpage
% \Urlmuskip=0mu plus 2mu\relax
\bibliographystyle{plain}
\bibliography{ref}
\clearpage
\appendix
\end{document}